\documentclass[fleqn,usenatbib]{mnras}

\usepackage{newtxtext,newtxmath}

\usepackage[T1]{fontenc}

\DeclareRobustCommand{\VAN}[3]{#2}
\let\VANthebibliography\thebibliography
\def\thebibliography{\DeclareRobustCommand{\VAN}[3]{##3}\VANthebibliography}

\usepackage{graphicx}	
\usepackage{amsmath}	

\usepackage{color} 
\usepackage{soul} 

\definecolor{GMgreen}{rgb}{0.1, 0.5, 0.05}

\title[ICL Assembly in \text{Horizon-AGN}]{Assembly of the Intracluster Light in the \textsc{Horizon-AGN} Simulation}

\author[Brown et al.]{
Harley J. Brown$^{1}$\thanks{E-mail: Harley.Brown@nottingham.ac.uk},
Garreth Martin$^{1}$,
Frazer R. Pearce$^{1}$,
Nina A. Hatch$^{1}$,
\newauthor
~Yannick M. Bah\'{e}$^{2}$,
and Yohan Dubois$^{3}$
\\
$^{1}$School of Physics \& Astronomy, University of Nottingham, University Park, Nottingham NG7 2RD, UK\\
$^{2}$Laboratory of Astrophysics, Institute of Physics, Ecole Polytechnique Fédérale de Lausanne (EPFL), Observatoire de Sauverny, Versoix 1290, Switzerland\\
$^{3}$Institut d’Astrophysique de Paris, CNRS, 98 bis blvd Arago,  Paris F-75014, France\\
}

\date{Accepted XXX. Received YYY; in original form ZZZ}

\pubyear{2024}

\begin{document}
\label{firstpage}
\pagerange{\pageref{firstpage}--\pageref{lastpage}}
\maketitle

\begin{abstract}
The diffuse stellar component of galaxy clusters made up of intergalactic stars is termed the intracluster light (ICL). Though there is a developing understanding of the mechanisms by which the ICL is formed, no strong consensus has yet been reached on which objects the stars of the ICL are primarily sourced from. We investigate the assembly of the ICL starting approximately $10$\,Gyr before $z=0$ in 11 galaxy clusters (halo masses between $\sim1\times 10^{14}$\,M\textsubscript{\sun} and $\sim7\times 10^{14}$\,M\textsubscript{\sun} at $z\approx0$) in the \textsc{Horizon-AGN} simulation. By tracking the stars of galaxies that fall into these clusters past cluster infall, we are able to link almost all of the $z\approx0$ ICL back to progenitor objects. Satellite stripping, mergers, and pre-processing are all found to make significant contributions to the ICL, but any contribution from in-situ star-formation directly into the ICL appears negligible. Even after compensating for resolution effects, we find that approximately $90$ per cent of the stacked ICL of the 11 clusters that is not pre-processed should come from galaxies infalling with stellar masses above $10^{9}$\,M\textsubscript{\sun}, with roughly half coming from infalling galaxies with stellar masses within half a dex of $10^{11}$\,M\textsubscript{\sun}. The fact that the ICL appears largely sourced from such massive objects suggests that the ICL assembly of any individual cluster may be principally stochastic. 
\end{abstract}




\section{Introduction}

The diffuse ensemble of stars that permeates the intergalactic space within a galaxy cluster is known as the intracluster light, or ICL. Said to have been first theorised and then discovered by \citet{zwicky_masses_1937, zwicky_coma_1951}, this low surface-brightness cluster component has been traced out to $\gtrsim1$\,Mpc from the cluster centre in stacked observational studies \citep{zhang_dark_2019}, and is expected to contribute significantly to the stellar mass budget of any cluster -- between $\sim5$ and $\sim50$ per cent (varying between clusters and with the exact definition used for the ICL; see \citealt{Mihos_review_2016}, \citealt{Contini_review_2021}, or \citealt{Montes_review_2022}).

It is currently thought that the ICL is formed through three primary channels: from the stellar detritus of massive galaxy mergers (e.g. \citealt{murante_importance_2007}, \citealt{contini_different_2018}), from stars stripped from satellite galaxies within the cluster by gravitational interactions (e.g. \citealt{willman_origin_2004}, \citealt{rudick_tidal_2009}), and also from the accreted intragroup light (IGL) of galaxy groups that previously merged with the cluster (e.g. \citealt{rudick_formation_2006}, \citealt{mihos_burrell_2017}), with this final formation channel commonly called ICL pre-processing. Though once considered potentially significant formation channels, total disruption of dwarf galaxies (e.g. \citealt{purcell_shredded_2007}) and in-situ formation of stars into the ICL (e.g. \citealt{puchwein_intracluster_2010}) are now thought to typically only minimally contribute to the ICL (\citealt{martel_fate_2012}, \citealt{melnick_intergalactic_2012}, \citealt{demaio_lost_2018}, and \citealt{gullieuszik_gasp_2020}; but also see \citealt{ahvasi_insitu_2024} and Bah\'{e} et al [in preparation] as well). 

Which or whether any of the three primary formation channels for the ICL is dominant is not yet fully understood, though there is a developing consensus in support of the two-phase formation scenario (\citealt{Kluge_twophaseform_2020}, \citealt{golden-marx_characterizing_2023}) in which stripping in general contributes the most by $z\sim0$, having potentially superseded the overall contribution from mergers sometime around $z\sim1$ (see \citealt{joo_intracluster_2023}, \citealt{Jimenez_supersede_2024} and references therein). Numerous recent studies -- both theoretical (e.g. \citealt{Tang_ICLfromSats_2023}, \citealt{contini_moreimptime_2024}) and observational (e.g. \citealt{montes_intracluster_2018}, \citealt{yoo_intracluster_2021}) -- support the idea of stripping being the generally dominant formation channel at $z\approx0$, especially for lower mass haloes \citep{contini_mwhalo_2024}, whereas with more massive haloes pre-processing is expected to play an increasingly significant role (\citealt{ragusa_does_2023}, \citealt{contini_moreimptime_2024}, and \citealt{chun_formation_2024}). For any specific individual cluster, however, the dominant source as well as other properties of the ICL will be intimately linked with the dynamical state of that specific cluster \citep{chun_formation_2023}.

It has been shown that ICL can provide remarkable insight into the accretion history of a cluster, with the ICL serving as a fossil record for cluster dynamics \citep{Montes_review_2022}. Furthermore, as the constituent stars of the ICL are collision-less, with their motions governed primarily by the overall cluster potential rather than that of individual galaxies, it is expected that these stars will behave much like dark matter and so should be distributed similarly within the cluster. It has thus been proposed that the ICL might be used as a visible dark matter tracer (e.g. \citealt{Montes_lumtrace_2019}, \citealt{alonsoasensio_intracluster_2020}, and \citealt{yoo_spatial_2024}) and also that it might reveal key features of the cluster potential such as the splash-back radius (e.g. \citealt{deason_stellar_2020}, \citealt{gonzalez_discovery_2021}).

Advances in recent years, both in technique and instrumentation, have pushed observational surface brightness limits to sufficiently low levels to enable revolutionary new ICL studies (see \citealt{mihos_deep_2019} for a recent brief review) such as the first ICL study using \textit{JWST} data by \citet{montes_jwst_2022}, and that by \citet{kluge_euclid_2024} using \textit{Euclid} Early Release Observations to study the ICL of the Perseus cluster. However, these observational studies require theoretical counterparts so that their results might be interpreted and contextualised, necessitating investigations into the predictions of numerical simulations concerning the ICL.

Though there is a developing consensus on the dominant formation channel of the ICL, a consensus has not yet been reached on the objects from which the stars of the ICL are primarily sourced. Theoretical studies generally agree on more massive galaxies as the more likely source for the majority of ICL stars by $z\sim0$ (though not unanimously), whilst observational studies remain more strongly divided. 

Numerical simulations of galaxy clusters can be classified into a number of distinct types which adopt different methods (see \citealt{somerville_physical_2015} for a review), including semi-analytical models (SAMs) and numerical hydrodynamic simulations. SAMs couple a simplified, analytical description of baryonic physics post hoc onto an existing N-body dark matter simulation, whilst hydrodynamical simulations instead numerically solve equations for gravity, thermodynamics, and hydrodynamics, for dark matter, gas, and stars simultaneously. Though relatively computationally inexpensive, baryons are not explicitly modelled in SAMs and so the physical processes responsible for the formation of the ICL are included only through approximate, simplified models. Hydrodynamical models (see \citealt{vogelsberger_cosmological_2020} for a review) are considerably more computationally expensive than SAMs, but allow direct and self consistent modelling of the physical processes that produce the ICL. However, limited computational resources force these simulations to compromise between total simulation volume and the resolution at which the physics can be explicitly simulated. Realistically modelling ICL formation requires simulation volumes large enough to provide cosmological context but this comes at the cost of a decreased resolution, which reduces the accuracy of the simulation at the small scale and may lead to non-physical numerical resolution effects.

Theoretical studies using SAMs have previously found that the main contributors of stars to the ICL by $z=0$ should be galaxies with stellar masses $\gtrsim 10^{10.5}\,\textrm{M\textsubscript{\sun}}$ \citep{contini_formation_2014, contini_theoretical_2019} and that the ICL can be heavily influenced by a small number of massive progenitors -- between $\sim1$ and $\sim10$ per cent the mass of the BCG \citep{cooper_surface_2015}. \citet{harris_quantifying_2017} re-simulated a Fornax-like cluster (halo mass $\sim 4\times 10^{13}\,\textrm{M\textsubscript{\sun}}$) from a dark matter only N-body simulation by replacing haloes infalling after $z=1.65$ with full galaxy models, and found $>60$ per cent of the $z=0$ ICL of this cluster to have been sourced from just two massive objects (stellar masses $\sim5\times 10^{10}\,\textrm{M\textsubscript{\sun}}$). By employing a similar ``galaxy replacement technique'' \citet{chun_formation_2023} studied 6 clusters (virial masses of order $10^{14}$\,M\textsubscript{\sun}) and found the ICL in all but the most relaxed cluster to be dominated by stars from galaxies with infall stellar masses $\gtrsim 10^{10}\,\textrm{M\textsubscript{\sun}}$. Continuing this work, \citet{chun_formation_2024} reported the typical progenitors of the IGL/ICL in 84 simulated groups and clusters ($13.6<\mathrm{log}_{10}(M_{200}[\mathrm{M_{\sun}}])<14.8$) to be galaxies with stellar masses between $10^{10}$\,M\textsubscript{\sun} and $10^{11}$\,M\textsubscript{\sun} on cluster infall, though also noted that the ICL of particularly massive or unrelaxed clusters could have significant contributions from galaxies with infall stellar masses $>10^{11}$\,M\textsubscript{\sun} as well. \citet{ahvazi_progenitors_2024} conducted a study on 39 groups and clusters (virial masses between $5 \times 10^{12}\,\textrm{M\textsubscript{\sun}}$ and $2 \times 10^{14}\,\textrm{M\textsubscript{\sun}}$) in the \textsc{TNG50} cosmological (magneto)hydrodynamical simulation \citep{nelson_illustristng_2021} and found that half the IGL/ICL across all the considered systems was brought in by galaxies with stellar masses between $10^{10}\,\textrm{M\textsubscript{\sun}}$ and $10^{11}\,\textrm{M\textsubscript{\sun}}$, with the ICL of some systems almost entirely originating from objects as or more massive than the Milky Way. 

Contrarily, when \citet{Tang_ICLfromSats_2023} analysed mock images of massive clusters at redshifts between $\sim0.1$ and $\sim1$ from the \textsc{TNG100} simulation \citep{nelson_illustristng_2021} they found the ICL to lie closest in the age-metallicity plane to satellites with stellar masses between $10^{8}\,\textrm{M\textsubscript{\sun}}$ and $10^{10}\,\textrm{M\textsubscript{\sun}}$. They consequently suggested that these intermediate mass galaxies were the main source of the ICL at these redshifts, as opposed to more massive galaxies. When considering the tension between their results and those of other theoretical studies, however, they also note that the methods they employed to extract galaxy stellar masses in these mock observations (from \citealt{tang_importance_2021}) could yield lower masses than would be found by traditional substructure extraction algorithms. 

Many observational studies support the picture generally presented by simulations -- that the stars of the ICL are primarily sourced from galaxies with stellar masses equal to or greater than approximately $10^{10}$\,M\textsubscript{\sun}. For example, \citet{montes_buildup_2021} suggested -- based on considerations of colour profiles -- that the ICL of A85 ($z\approx0.05$, $M_{200}\approx1.7\times10^{15}$\,M\textsubscript{$\sun$}) was built up mainly by the stripping of satellites with stellar masses of order $10^{10}\,\textrm{M\textsubscript{\sun}}$. Similarly, \citet{montes_intracluster_2014, montes_intracluster_2018} found (also based on colours) that the ICL of several massive ($M_{200}>10^{14.3}$\,M\textsubscript{$\sun$}) clusters at $0.3 < z < 0.6$ appeared to primarily come from Milky-Way-like objects, and \citet{demaio_lost_2018} found (on the basis of colour matching) that $75$ per cent of the IGL/ICL in the groups and clusters ($3\times10^{13}\lesssim M_{500}/$\,M\textsubscript{$\sun$}$ \lesssim 9\times10^{14}$) they considered at $0.29 < z < 0.89$ should have originated from galaxies with stellar mass $> 10^{10.4}\,\textrm{M\textsubscript{\sun}}$. 

Conversely, \citet{morishita_characterizing_2017} found the ICL of six \textit{Hubble} Frontier Field clusters ($0.3<z<0.6$, $M_{500}\gtrsim10^{15}$\,M\textsubscript{$\sun$}) to be dominated by moderately old stars ($\sim 1$ to $3$\,Gyr), and found the colours of these stars to be more consistent with these having been stripped mainly from cluster galaxies with stellar masses less than $\sim10^{9.5}\,\textrm{M\textsubscript{\sun}}$ since $z\sim1$, instead of more massive galaxies. Likewise, \citet{gu_spectroscopic_2020} studied portions of the ICL in the Coma cluster ($z\approx0.024$, $M_{200}\approx5.1\times10^{14}$\,M\textsubscript{$\sun$}: \citealt{gavazzi_weak_2009}), unearthing a very old and metal-poor stellar population -- similar to that found by \citet{williams_metallicity_2007} for intracluster stars in the Virgo cluster ($z\approx0.004$, $M_{200}\approx1.4\times10^{14}$\,M\textsubscript{$\sun$}: \citealt{urban_x-ray_2011}) -- and suggested that these stars may have primarily originated from the accretion of low-mass galaxies (stellar masses less than $\sim3\times 10^9$\,M\textsubscript{\sun}). However, the authors of both these studies note that their analyses consider only typical galaxy colours and metallicities, and invoke the presence of strong colour gradients and lower outskirt metalliticites in more massive galaxies -- and that the stars in these galaxies at large galactocentric radii should be more easily stripped and so make up a disproportionate fraction of the stars these galaxies contribute to the ICL -- as a potential explanation for the tension between their results and the emerging consensus from theoretical studies. 

The tension that remains between predictions for the main progenitors of the ICL -- both between theoretical and observational studies, and between different studies within each of these categories -- merits further investigation. A small number of prior studies have already investigated the primary progenitors of ICL stars as predicted by hydrodynamical simulations (e.g. \citealt{ahvazi_progenitors_2024}). However, their conclusions are likely, at least in part, dependent on the specific physical models used by each different simulation code and there is therefore worth in investigating this area with additional simulations. 

In this paper we present a new study of the origin and formation of the ICL using the \textsc{Horizon-AGN} simulation \citep{dubois_dancing_2014}. We investigate the assembly of the ICL starting approximately $10$\,Gyr before $z=0$ in 11 clusters with dark matter halo masses between $\sim1\times10^{14}\,$M\textsubscript{\sun} and $\sim7\times10^{14}\,$M\textsubscript{\sun} at $z\approx0$. By tracking stars associated with galaxies that fall into these clusters during this time, we are able to link much of the $z\approx0$ ICL with progenitor galaxies and so investigate the relative contributions to the ICL of galaxies with differing stellar masses preceding cluster infall. Additionally, by extrapolating our results for intermediate and high stellar mass progenitor objects ($\gtrsim10^{9}\,\mathrm{M_{\sun}}$) down to lower masses to compensate for resolution effects, we estimate the range of galaxy stellar masses responsible for the bulk of the ICL in a typical cluster.

This paper is organised as follows: In Section~\ref{Methods} we present the details of the \textsc{Horizon-AGN} simulation (\ref{Methods:HAGN}) as well as those of the \textsc{AdaptaHOP} structure finder and \textsc{Treemaker} merger tree builder, and then describe how we employ these for our investigation into the assembly of the ICL (\ref{Methods:AdaptaHOP_trees_and_my_code}). In Section~\ref{results} we describe and discuss the results of our investigation -- first on the stellar mass budgets of the 11 clusters and what fraction of ICL stars could formerly be found in galaxies within the clusters (\ref{ssec:vp_plots}), then on the contributions of progenitor objects with differing stellar masses on cluster infall to the ICL (\ref{sec:main_results}). We conclude by summarising the main results of our investigation in Section~\ref{conclusions}. 
Throughout this paper, we assume a $\Lambda$CDM cosmology with $h \equiv H_{0} / 100 \text{\,km\,s\textsuperscript{$-1$}\,Mpc\textsuperscript{$-1$}} = 0.704$. Unless stated otherwise, distances are given in proper rather than co-moving units. 

\section{Methods}\label{Methods}

\subsection{\textsc{Horizon-AGN}}\label{Methods:HAGN}

The full details of the \textsc{Horizon-AGN} simulation can be found in \citet{dubois_dancing_2014, dubois_horizon-agn_2016}, and are reproduced only in brief here. 

The \textsc{Horizon-AGN} simulation is a cosmological-volume hydrodynamical simulation which employs $1024^3$ dark matter (DM) particles (each with mass $8 \times 10^{7}\,$M\textsubscript{\sun}) in a cubic volume of side-length $100\,h^{-1}\,$Mpc (co-moving) with periodic boundary conditions. A standard $\Lambda$CDM cosmological model \citep{peebles_cosmological_2003} is used with total matter density $\Omega_{\mathrm{m}}=0.272$, baryon density $\Omega_\mathrm{b}=0.045$, dark energy density $\Omega_\Lambda=0.728$, dark matter power spectrum amplitude $\sigma_{8}=0.81$, Hubble constant $H_0 = 70.4$\,km\,s\textsuperscript{$-1$}\,Mpc\textsuperscript{$-1$}, and spectral index $n_{\mathrm{s}}=0.967$, compatible with the \textit{Seven-Year Wilkinson Microwave Anisotropy Probe} cosmology \citep{komatsu_seven-year_2011}, and also that of the \citet{planck_collaboration_planck_2014} within $\sim10$ per cent relative variation. The simulation utilises \textsc{Ramses} \citep{teyssier_cosmological_2002} -- an adaptive mesh refinement Eulerian hydrodynamics code -- with an initially uniform $1024^3$ cell gas grid, refined according to a quasi-Lagrangian criterion (based on either the total baryonic or DM mass in a cell exceeding eight times the mass of a DM particle), down to a minimum cell size of $\Delta x = 1$\,kpc after seven levels of refinement. 

Gas heating from a uniform UV background begins after $z=10$, following \citet{haardt_radiative_1996}, and gas is allowed to cool to $10^4$\,K via H and He collisions with a contribution from metals using a \citet{sutherland_cooling_1993} model. Feedback from active galactic nuclei is a key process for regulating the stellar mass content of massive galaxies \citep{dubois_horizon-agn_2016}, and active galactic nuclei are modelled in \textsc{Horizon-AGN} by the accretion of gas onto supermassive black holes following a Bondi-Hoyle-Lyttleton accretion rate \citep{bondi_spherically_1952} capped at the Eddington rate, with switching between jet (``radio'') and heating (``quasar'') feedback modes according to accretion rate \citep{dubois_self-regulated_2012}. Star formation is modelled using a Schmidt law, with $2$ per cent star formation efficiency \citep{kennicutt_jr_global_1998}, and is only allowed in regions with gas number density exceeding $0.1$\,H\,cm\textsuperscript{$-3$}. Feedback is included from Type Ia and Type II supernovae, as well as stellar winds, through mass, energy, and metal release. The star particles (i.e. the resolution elements of the ICL) have an initial mass resolution of approximately $2\times10^6$\,M\textsubscript{\sun}. 

The \textsc{Horizon-AGN} simulation was not calibrated to the local Universe, except for choosing blackhole feedback parameters. Despite this, \citet{kaviraj_horizon-agn_2017} reported that the simulation is in generally good agreement with observations (using observational data from $0<z<6$) concerning e.g. predicted evolution of luminosity functions, stellar mass functions, the star formation main sequence, and the cosmic star formation history. A slight overabundance of galaxies with stellar masses $\lesssim10^{10.5}$\,M\textsubscript{\sun} at all epochs is noted compared to observations, though agreement can be achieved by considering observational uncertainties due to cosmic variance. 

\subsection{Tracking galaxies}\label{Methods:AdaptaHOP_trees_and_my_code}

\subsubsection{\textsc{AdaptaHOP} and \textsc{Treemaker}}\label{Methods:AdaptaHOP_Trees}

For identifying structures of both DM particles (i.e. haloes) and star particles (i.e. galaxies), the \textsc{AdaptaHOP} structure finder \citep{aubert_origin_2004} is employed (and specifically the version updated by \citealt{tweed_building_2009} for building merger trees), the details of which are reproduced only in brief here.

Particles are grouped into structures on the basis of local particle density, calculated using a nearest neighbours approach with 20 neighbour particles. No unbinding procedure is employed when creating these particle groups. Stellar structures and substructures are assembled from the groups created by linking star particles with densities $>178\times$ the total matter density (of the entire simulated volume) according to their closest local density maxima. Saddle points in the density field between these groups are then used to link these groups together into stellar structures (i.e. galaxies), which are then subdivided hierarchically on the basis of increasing density to identify substructure. The distance used for force softening is $\sim2$\,kpc (hence substructures smaller than this are considered irrelevant), and stellar structures with $\leq50$ star particles are neglected. As the star particles have a mass of $\sim2\times10^6$\,M\textsubscript{\sun}, this gives a minimum (detectable) galaxy stellar mass of $\sim 10^{8}$\,M\textsubscript{\sun}. We address the potential contribution to the ICL from galaxies less massive than this detection threshold in Section~\ref{ssec:vp_plots}.

Essentially the same procedure is used to identify DM structures (and substructures), though the density threshold used is instead $80\times$ the total matter density, and the minimum membership threshold for a structure to not be neglected is raised to $100$ particles. Galaxies and DM haloes are identified independently and only afterwards are galaxies linked to host haloes. The main galaxy of a halo is defined to be the most massive galaxy within $0.1\times r_{178}$ of the halo centre (where $r_{178}$ is the radius from the halo centre within which the average DM density is $178\times$ the critical density).

The \textsc{Treemaker} algorithm (originally developed by \citealt{hatton_galics-_2003}; see also \citealt{tweed_building_2009}) is used to generate merger trees for the galaxies identified by \textsc{AdaptaHOP}. In brief, each structure in each simulation snapshot is (if possible) linked to a main descendent structure in the next snapshot and a main progenitor in the previous snapshot using a merit function which maximises the fraction of particles shared by the would-be linked structures. Between $z\approx3$ and $z\approx0$ the mergers trees we use in this study were built using a time resolution of $\sim0.05$\,Gyr. 

For this study the galaxy considered to be the BCG of each cluster is selected at $z\approx0$ ($z=0.0556$) as the most massive galaxy within $0.1\times r_{178}$ of the cluster halo centre. Prior to $z\approx0$ the galaxy classified as the BCG of the cluster in each snapshot is the main progenitor of the $z\approx0$ BCG. We place the border of each cluster in each snapshot at a distance away from the BCG equal to $r_{178}$ of the cluster halo (see Section~\ref{Methods:mycode}) and classify all galaxies within this border besides the BCG as satellite galaxies. 

\subsubsection{Tagging and tracking the stars of infalling galaxies}\label{Methods:mycode}

For this investigation we select 11 clusters from the \textsc{Horizon-AGN} simulation, with $z\approx0$ ($z=0.0556$) DM halo masses (within $r_{178}$ i.e. $M_{178}$) in the range $\sim1 \times 10^{14}$ to $\sim7 \times 10^{14}$\,M\textsubscript{\sun} and investigate their ICL assembly. We restrict our analysis to cubes, side length $8$\,Mpc, centred on the BCG of each cluster. 
For each cluster, we perform our analysis using a coarse time resolution of $\sim1$\,Gyr, down to $z\approx0$ ($z=0.0556$) from as far back as when a main progenitor of the cluster BCG is first identified (which is always before $z\sim1.5$ and almost always before $z\sim2$) or $z\sim3$ -- whichever is later. 
As later described in Section~\ref{ssec:vp_plots}, we never find more than $\sim0.1$ per cent of the $z\approx0$ ICL stars of any of the 11 clusters to have assembled before we begin to monitor ICL assembly. This is in good agreement with previous theoretical studies (e.g. \citealt{willman_origin_2004} and \citealt{contini_moreimptime_2024}) which typically found the vast majority of the $z=0$ ICL mass of similarly massive clusters to have still been contained in galaxies at $z\sim1$. 

We place the border of each cluster at $r_{178}$ and consider galaxies to have become satellites of a cluster once they pass $r_{178}$. Rather than use the values provided by \textsc{AdaptaHOP} directly, for each cluster we instead fit a monotonic cubic spline to the $r_{178}$ values of each cluster from \textsc{AdaptaHOP} as a function of look-back time (using snapshots with $\sim1$\,Gyr coarse spacing), allowing us to account for and smooth out any temporary and nonphysical fluctuations in cluster size (such as those caused by cluster mergers). 

To identify the progenitor objects of the $z\approx0$ ICL, we track the star particles of galaxies falling into the 11 clusters starting from the coarse snapshot immediately proceeding galaxy infall down to $z\approx0$. Infalling galaxies are identified as galaxies at $r>r_{178}$ in any coarse snapshot whose main descendant in the next coarse snapshot is at $r<r_{178}$. This infaller galaxy identification is done chronologically, so galaxies which have previously fallen into the cluster but whose orbit apocentre still lies outside $r_{178}$ can be identified (and repeat-counting avoided) by confirming that each candidate infaller galaxy does not have a previously identified infaller as a main progenitor. When a galaxy infalling for the first time is identified, all of its stars in the coarse snapshot immediately proceeding infall are tagged as associated with that progenitor, and the galaxy then followed through all remaining coarse snapshots using the merger trees. 

Any new star particles born in the descendent of an infaller are also tagged as associated with that infaller, provided that the infaller galaxy is still part of the main progenitor branch of the descendant galaxy those star particles are born in. This restriction is imposed to avoid double-counting (so these new star particles are only linked to a single progenitor rather than potentially several progenitors that merged after infall). 
We find these ``newborn'' star particles -- those born during or after galaxy infall -- to make up $\sim4$ per cent of all $z\approx0$ ICL stars across all 11 clusters (and $\sim6$ per cent of all not pre-processed ICL stars), so include them in our analysis for completeness, attributing them as contributed to the ICL by the progenitor object they are tagged to and equivalent to the stars of the same galaxy that formed long before cluster infall. However, by repeating our analysis while ignoring these newborn stars we confirm that whether or not they are included does not meaningfully alter our conclusions. 

We use the merger trees generated by \textsc{Treemaker} to follow galaxies past cluster infall, but the extreme environment of a cluster makes errors in linking progenitor and descendent objects somewhat common, leading to some missing or incorrect links. Whenever we identify a break in the merger tree of a tracked galaxy or find that no star particles are shared between a tracked galaxy and its supposed descendent in the next coarse snapshot, and another galaxy can also be found in the next coarse snapshot which contains $>50$ per cent of the tracked galaxy's star particles, we overrule \textsc{Treemaker} and classify the latter galaxy as the tracked galaxy's proper descendent. Of the $\sim4000$ infalling galaxies tracked, we modify the merger-trees of $\sim15$ per cent after initial infall -- most commonly to patch breaks while galaxies are partway through merging with the BCG.

\subsubsection{Identifying the ICL}\label{Methods:example_cluster_plots}

\begin{table*}
    \centering
        \caption{Various properties of the 11 simulated clusters. The assembly redshifts are those when a main progenitor of the cluster halo first has a total DM mass equal to or exceeding 50 per cent of that of the $z\approx0$ cluster. The DM and stellar masses are those within $r_{178}$ at $z\approx0$ and include substructure, and $f_{\textrm{BCG}}$, $f_{\textrm{Sat}}$, and $f_{\textrm{ICL}}$ are the fractions of this stellar mass in the BCG, satellite galaxies, and ICL respectively and are also depicted in Figure~\ref{fig:stel_mass_vp}. The ICL sub-fractions given in the last three columns are also depicted in Figure~\ref{fig:ICL_orig_vp}.}
    \begin{tabular}{ |c|c|c|c|c|c|c|c|c|c|c| }
         \hline
         Cluster & DM Halo Mass & Stellar Mass & Assembly & $f_{\textrm{BCG}}$ & $f_{\textrm{Sat}}$ & $f_{\textrm{ICL}}$ & ICL Fraction Directly & ICL Fraction & ICL Fraction From\\
         ID & [$10^{14}$\,M\textsubscript{\sun}] & [$10^{12}$\,M\textsubscript{\sun}]& Redshift & & & & From Satellites & Previously in BCG & Pre-Processing\\
         \hline
         1 & 1.44 & 4.74 & 0.415 & 0.304 & 0.566 & 0.130 & 0.523 & 0.339 & 0.138 \\
         9 & 0.973 & 2.84 & 1.24 & 0.506 & 0.352 & 0.142 & 0.442 & 0.476 & 0.0818\\
         13 & 6.86 & 19.6 & 0.122 & 0.132 & 0.740 & 0.129 & 0.446 & 0.115 & 0.438\\
         19 & 5.32 & 14.4 & 0.594 & 0.338 & 0.495 & 0.168 & 0.439 & 0.394 & 0.167\\
         46 & 1.17 & 3.54 & 0.566 & 0.336 & 0.549 & 0.115 & 0.511 & 0.368 & 0.121\\
         48 & 1.03 & 3.28 & 0.745& 0.353 & 0.518 & 0.129 & 0.435 & 0.423 & 0.142\\
         49 & 3.09 & 8.66 & 1.07 & 0.346 & 0.514 & 0.139 & 0.522 & 0.359 & 0.119\\
         71 & 1.34 & 4.78 & 0.369 & 0.316 & 0.572 & 0.112 & 0.437 & 0.298 & 0.264\\
         132 & 1.20 & 4.08 & 0.270 & 0.310 & 0.588 & 0.101 & 0.409 & 0.283 & 0.308\\
         174 & 2.09 & 6.17 & 0.233 & 0.248 & 0.623 & 0.128 & 0.584 & 0.190 & 0.226\\
         183 & 1.08 & 3.64 & 0.332 & 0.348 & 0.540 & 0.112 & 0.428 & 0.399 & 0.173\\
        \hline
    \end{tabular}
    \label{tab:cluster_props}
\end{table*}

We identify the ICL using the output from \textsc{AdaptaHOP}. Specifically, any star particles in the clusters not classed by \textsc{AdaptaHOP} as part of a stellar structure are considered to be part of the ICL (and so the definition for the ICL employed in this study is one based on instantaneous three-dimensional density). The ICL fractions (i.e. fraction of cluster stellar mass attributed to the ICL rather than cluster galaxies) according to this definition of the ICL for the 11 clusters at $z\approx0$ are shown in Table~\ref{tab:cluster_props}, along with a selection of other cluster properties. 

It bears acknowledging that different ICL identification methodologies (e.g. different structure finders) will disagree on the exact stellar content of the ICL, so our choice of structure finder will have some impact on our results. However, prior studies have shown broadly good agreement across a wide range of structure finder codes (for halo properties such as virial mass, bulk velocity, rotation velocity, and the presence and location of substructure: see \citealt{knebe_haloes_2011},  \citealt{onions_subhaloes_2012} and references therein), and \citet{brough_preparing_2024} demonstrated that \textsc{AdaptaHOP} yields ICL properties in good agreement with those recovered using other structure finder codes, such as \textsc{Subfind} \citep{dolag_substructures_2009} which includes an unbinding procedure. As such, we do not anticipate our choice of structure finder will significantly influence our results. We address the related issue of splitting the BCG and the ICL in Section~\ref{ssec:vp_plots}. 

\begin{figure*}
    \includegraphics[width=2.07\columnwidth]{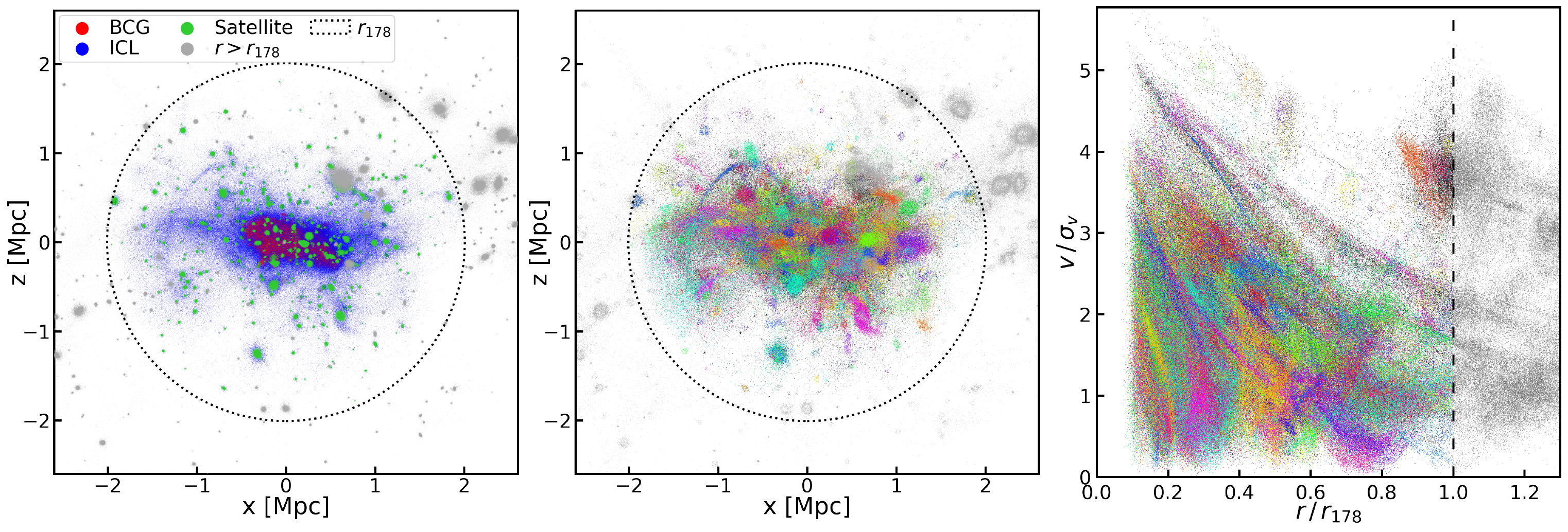}
    \caption{\textbf{Left}: Projected positions of all the stars in the vicinity of an example $z\approx0$ \textsc{Horizon-AGN} galaxy cluster (ID 19 in Table~\ref{tab:cluster_props}) using a projection depth of 8\,Mpc (centred on the BCG) and coloured according to whether a star is considered part of the BCG (red), a satellite galaxy (green), part of the ICL (blue), or is outside of the cluster (grey). \textbf{Middle}: Projected positions of only intergalactic stars for the same cluster as shown on the left. ICL stars associated with the same progenitor galaxy are shown in the same colour; those not associated with a progenitor infaller (e.g. pre-processed ICL) are shown in black. Intergalactic stars outside of the cluster (e.g. the IGL of infalling groups) are shown in grey. \textbf{Right}: Scatter plot of the speeds, $v$, of intergalactic stars (relative to the BCG and scaled by the 3D velocity dispersion of the cluster stars, $\sigma_{v}$) against distance from BCG centre, $r$, (scaled by $r_{178}$) for the same cluster and using the same colours as the middle panel.}
    \label{fig:intro_s&vps}
\end{figure*}

An example of one of the 11 clusters (ID 19 in Table~\ref{tab:cluster_props}) is shown in Figure~\ref{fig:intro_s&vps}; the left and centre panels show the projected positions for an arbitrary line-of-sight at $z\approx0$ of all stars and only ICL stars respectively. The stars shown in the left panel are coloured according to whether they were classified by \textsc{AdaptaHOP} as being part of the BCG (red), a satellite galaxy (green), or the ICL (blue), whilst those shown in the central panel are coloured according to associated progenitor. ICL star particles unable to be associated with a progenitor object (such as those which did not enter the cluster as part of a galaxy i.e. pre-processed ICL) are shown in black; otherwise the specific colours used do not relate to any physical properties of the stars or their progenitor. The right panel in Figure~\ref{fig:intro_s&vps} is a velocity dispersion plot (speed relative to the BCG plotted against distance from BCG centre, normalised by the 3D velocity dispersion of all cluster star particles and $r_{178}$ respectively) for the same ICL star particles as shown in the centre panel of Figure~\ref{fig:intro_s&vps} and employing the same scheme for colours. Comparison between the centre and right panels shows correspondence between features in position and velocity space, as expected for ICL stars recently sourced from the same infalling galaxy.

\subsubsection{Uncertainty estimation}\label{Methods:bootstrapping}

Throughout the remainder of this paper all indicated uncertainties are estimated using bootstrapping. Specifically, we produce 5000 bootstrap resamples of our population of $\sim4000$ tracked infaller galaxies and their relative contributions to the ICL (sampled with replacement to produce infaller galaxy resamples of equal size to the original population), and repeat our analysis on each of these resamples. The indicated uncertainties are found using the 16\textsuperscript{th} and 84\textsuperscript{th} percentiles from these repeat analyses.

\section{Results}\label{results}

\subsection{Stellar mass budget and ICL pre-processing} \label{ssec:vp_plots}

\begin{figure}
	\includegraphics[width=\columnwidth]{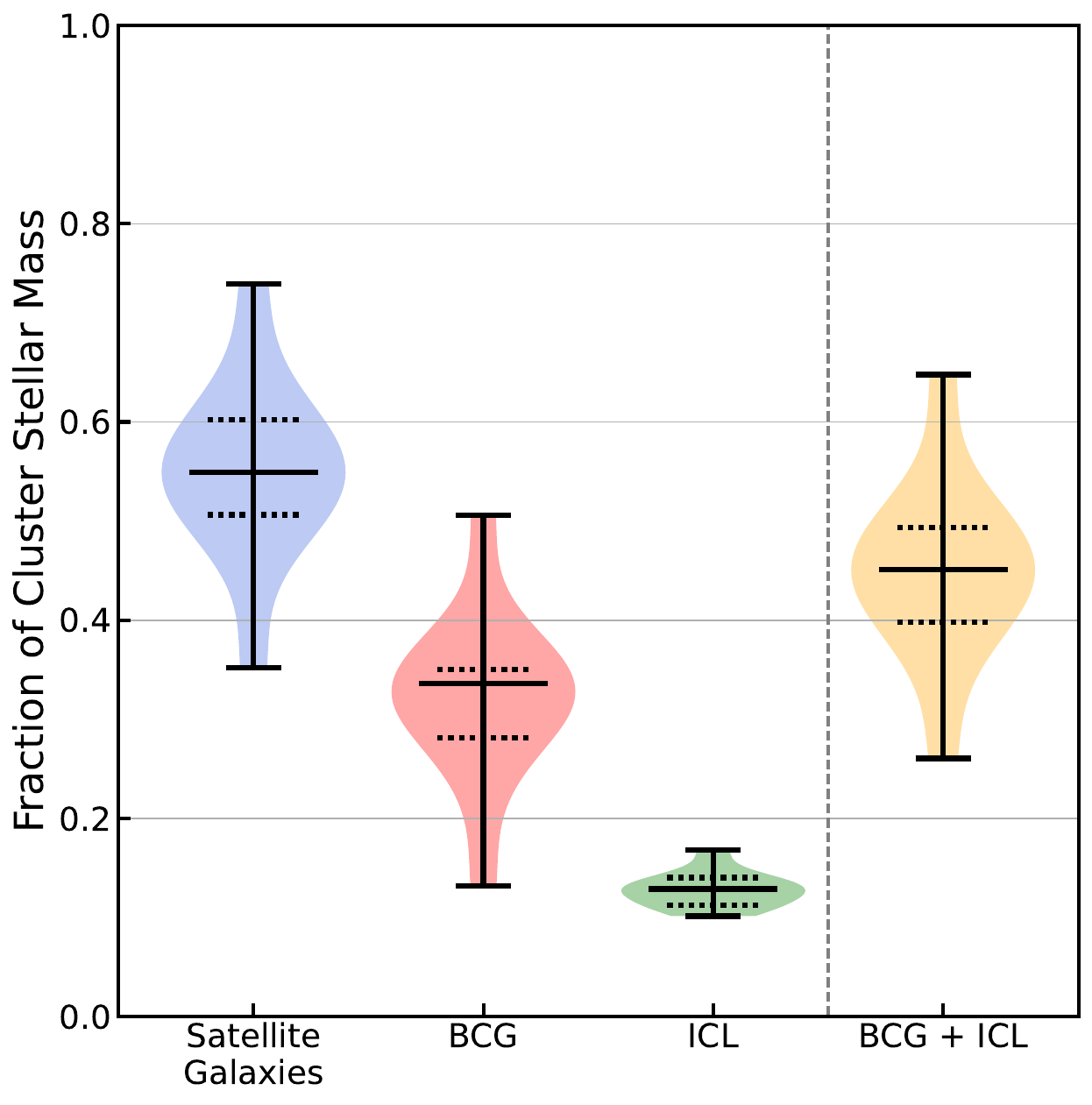}
    \caption{The $z\approx0$ stellar mass budgets of the 11 simulated clusters, divided between satellite galaxies, the BCG, and the ICL. Combined BCG\,+\,ICL stellar mass fractions also shown. Dotted lines indicate 16\textsuperscript{th} and 84\textsuperscript{th} percentiles, and solid lines the maximum, minimum, and median values.}
    \label{fig:stel_mass_vp}
\end{figure}

\begin{figure}
	\includegraphics[width=\columnwidth]{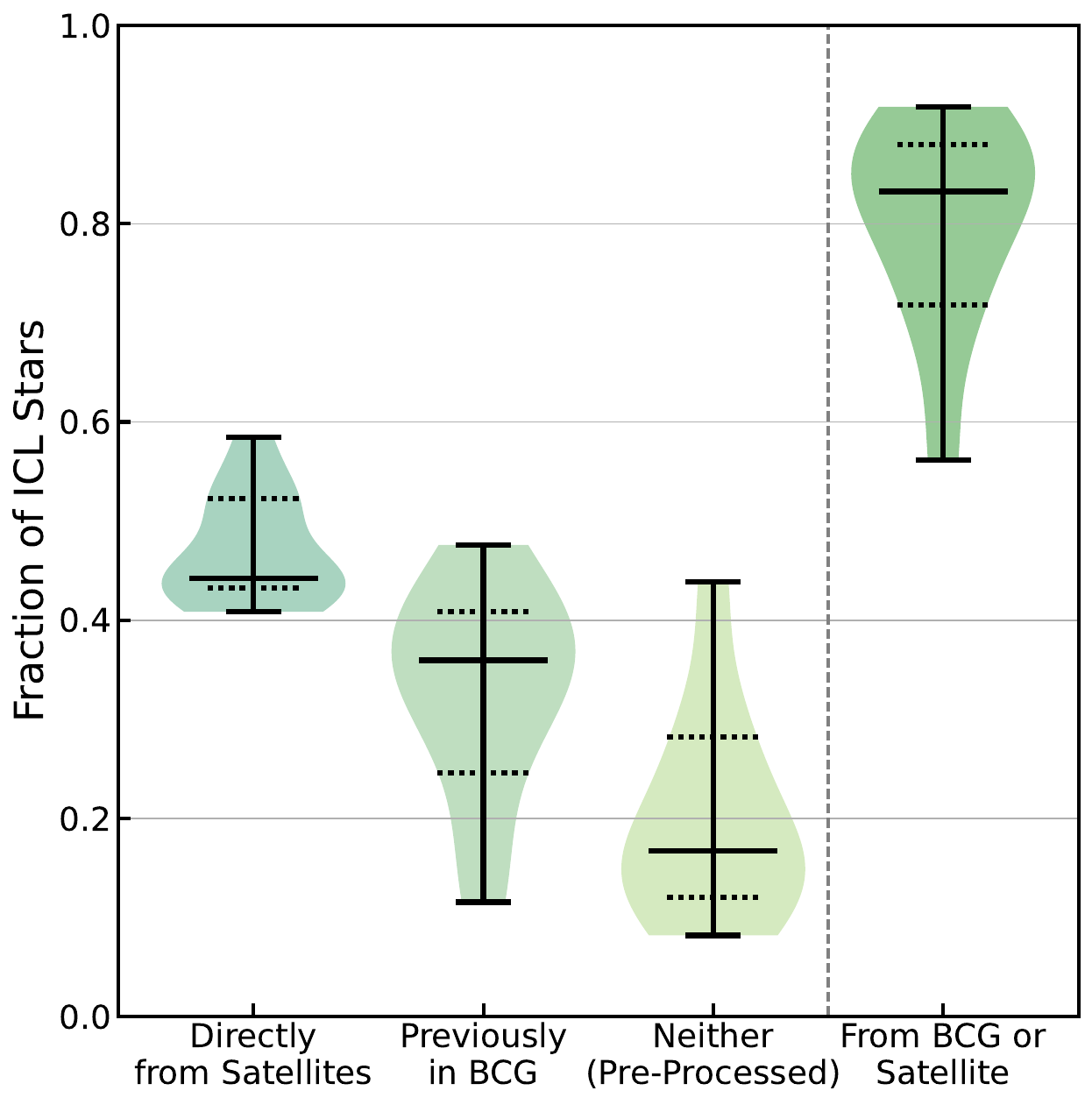}
    \caption{Fractions of $z\approx0$ ICL stars formerly part of satellite galaxies, previously part of the BCG, or that were never part of a cluster galaxy or an infalling galaxy (i.e. pre-processed ICL) in the 11 clusters. Combined fractions from either the BCG or directly from satellite galaxies (i.e. not pre-processed ICL) also shown. Dotted lines indicate 16\textsuperscript{th} and 84\textsuperscript{th} percentiles, and solid lines the maximum, minimum, and median values.}
    \label{fig:ICL_orig_vp}
\end{figure}

Figure~\ref{fig:stel_mass_vp} depicts how the $z\approx0$ stellar mass budgets of each of the 11 clusters are split between satellite galaxies, the BCG, and the ICL. The combined BCG\,+\,ICL fractions are also included. The median $z\approx0$ ICL (BCG\,+\,ICL) stellar mass fraction of the clusters is $13$ ($45$) per cent, the 16\textsuperscript{th} percentile is $11$ ($40$) per cent, and the 84\textsuperscript{th} percentile is $14$ ($49$) per cent. 

These ICL fractions fall within the wide range of values reported in the literature -- between $\sim5$ and $\sim50$ per cent (\citealt{Contini_review_2021}; see also \citealt{tang_investigation_2018} and \citealt{kluge_photometric_2021}) -- and agree fairly well with the findings of other theoretical studies employing similar ICL definitions, such as that by \citet{rudick_quantity_2011} who found ICL fractions between $\sim9$ and $\sim15$ per cent while employing an ICL definition based on instantaneous density in their investigation of a selection of simulated clusters with masses of order $\sim10^{14}$\,M\textsubscript{\sun}. These ICL fractions also broadly agree with many of those found for similarly massive clusters by studies which employed different ICL definitions -- both theoretical (e.g. $\sim9-14$ per cent: \citealt{ahvazi_progenitors_2024}) and observational (e.g. $\sim7-15$ per cent: \citealt{mihos_burrell_2017}) -- though some studies instead report much higher ICL fractions (e.g. $\gtrsim30$ per cent: \citealt{pillepich_first_2018}; $20-40$ per cent: \citealt{furnell_growth_2021}). 

A major driver of the large variation in ICL fractions found by both theoretical and observational ICL studies is the wide variety of different methods employed to measure the ICL \citep{brough_preparing_2024}. These different ICL measures differ most significantly in how they separate the ICL from the BCG, with large variations seen in the ICL fraction depending on where the border between the ICL and BCG is drawn. This perspective is supported by the broad agreement seen in BCG\,+\,ICL stellar fractions between studies -- generally around $\sim50$ per cent for clusters with masses of order $\sim10^{14}$\,M\textsubscript{\sun} (e.g. \citealt{pillepich_first_2018}, \citealt{brough_preparing_2024}, and \citealt{contreras-santos_characterising_2024}, in addition to this study) -- even when measured ICL fractions differ significantly. Consequently, rather than implement an uncertain border between them, some previous studies have opted to simply not separate the BCG from the ICL and consider the combined system only (e.g. \citealt{pillepich_first_2018}, \citealt{zhang_dark_2019}, and \citealt{chun_formation_2023}). In similar acknowledgement of this definition issue we repeat the remainder of our analysis while likewise not separating the BCG from the ICL, and where relevant present this alternative analysis alongside our primary analysis for comparison (and always refer to this combined system as BCG\,+\,ICL). This definition issue also has significant implications for any study of the radial dependence of ICL sourced from different formation channels, so we defer investigating such a dependence to future work. 

Figure~\ref{fig:ICL_orig_vp} depicts the fractions of $z\approx0$ ICL stars in the 11 clusters that were previously part of a satellite galaxy (but never part of the BCG), previously part of the BCG, or that have never been part of a cluster galaxy (i.e. were pre-processed). The combined fractions of ICL stars that were previously part of any cluster galaxy are also included. 

The median fraction of $z\approx0$ ICL stars formerly in satellite galaxies (and never part of the BCG) is $44$ per cent (16\textsuperscript{th} percentile $43$ per cent and 84\textsuperscript{th} percentile $52$ per cent), and the median fraction that were formerly part of the BCG is $36$ per cent (16\textsuperscript{th} percentile $26$ per cent and 84\textsuperscript{th} percentile $41$ per cent). That never less than $40$ per cent of the $z\approx0$ ICL stars in any of the 11 clusters were liberated directly from satellite galaxies within the cluster (and had never been part of the BCG) indicates stripping must play a significant role in ICL assembly. Likewise, that on average more than a third of $z\approx0$ ICL stars were formerly part of the BCG of their cluster implies violent BCG mergers can play a similarly significant role. We revisit the importance of BCG mergers for ICL assembly in Section~\ref{ssec:strip_frac_plot}. 

When \citet{montenegro-taborda_growth_2023} investigated the ICL assembly of $M_{200}\geq10^{14}$\,M\textsubscript{$\sun$} clusters in the \textsc{TNG300} simulation \citep{nelson_illustristng_2021}, they found the mean fraction of ICL stars (at $z=0$) that had been stripped from surviving satellites to be $38.1$ per cent. 
The value we find for the median fraction of ICL stars directly stripped from satellite galaxies ($\sim44$ per cent) is somewhat larger than this, as would be expected since (unlike \citealt{montenegro-taborda_growth_2023}) we do not exclude ICL stars stripped from satellite galaxies that later merge with the BCG or otherwise do not survive until $z=0$.
\citet{chun_formation_2024} reported a median value of $\sim50$ per cent for the fraction of the $z=0$ ICL stars of $14<\textrm{log}(M_{200}/$M\textsubscript{$\sun$}$)<14.4$ clusters that could be linked to the merger tree of the cluster BCG. As this value combines together the ICL contributions from BCG mergers and from the tidal stripping of galaxies prior to merging with the BCG, it is unsurprising that this value is larger than the median fraction of the ICL we find to have previously been part of the BCG ($\sim36$ per cent). We thus do not consider our results in tension with those of \citet{montenegro-taborda_growth_2023} or \citet{chun_formation_2024}. 

Using SAMs \citet{contini_moreimptime_2024} obtained a median $z=0$ ICL contribution from mergers of only $\sim10$ per cent for clusters with halo masses between $\sim10^{14}$\,M\textsubscript{\sun} and $\sim10^{14.5}$\,M\textsubscript{\sun}. This low fraction does not appear compatible with the median fraction of ICL stars we find to have formerly been part of the BCG ($\sim36$ per cent), as mergers should be the primary mechanism for liberating these stars. We speculate that this disagreement may stem from the simple prescription employed for mergers by the utilized SAM, which does not allow BCG stars to be added to the ICL (instead just moving $20$ per cent of the stellar mass of a merging satellite into the ICL component when a BCG ``merger'' occurs). The aforementioned issue of studies which employ differing methodologies for separating the BCG and the ICL determining significantly differing properties for the ICL may also play some part in this apparent incompatibility.

We note that not all of the $z\approx0$ ICL stars identified by \textsc{AdaptaHOP} as previously being part of a cluster galaxy actually entered that cluster as part of a galaxy. For example, $\sim20$ per cent of the $z\approx0$ ICL stars across all 11 clusters classed by \textsc{AdaptaHOP} as formerly being in satellite galaxies could not be linked to a progenitor infaller galaxy (as they neither entered the cluster as part of that infaller nor were they formed in its descendent). These stars would often appear distinct from the rest of their supposed galaxy in phase-space, and were frequently only classified as part of a cluster galaxy for a single coarse snapshot, having originally entered the cluster while not classed by \textsc{AdaptaHOP} as part of a galaxy (i.e. as pre-processed ICL). We suspect the transient galaxy membership of these stars to be a result of \textsc{AdaptaHOP} not employing any kind of unbinding procedure. The median fraction (across the 11 clusters) of ICL stars classed as formerly being part of satellite galaxies, the BCG, or either that could not be linked to a progenitor infaller galaxy were $19$, $27$, and $23$ per cent respectively. As these stars could not be associated with a progenitor infaller, they are not considered (along with all other pre-processed ICL) in our later analysis of the differing contributions of infalling galaxies of varying masses to the ICL. 

As a further consequence of \textsc{AdaptaHOP} not employing an unbinding procedure, we note that any ICL stars on radial orbits that are coincidentally caught within the BCG at the time of a coarse snapshot would be classed by \textsc{AdaptaHOP} as part of the BCG, and hence be added to the category referred to as ``previously in BCG'' in Figure~\ref{fig:ICL_orig_vp}. As a result, we caution that this category is not an ideal proxy for the ICL contribution from violent mergers with the BCG and should instead be interpreted as an upper limit for the contribution from this channel.


Among the 11 clusters, the median fraction of $z\approx0$ ICL stars that have never been classed as part of a cluster galaxy is $16$ per cent (16\textsuperscript{th} percentile $12$ per cent and 84\textsuperscript{th} percentile $28$ per cent). This is in middling agreement with the pre-processed ICL fractions found (using SAMs) by \citet{contini_moreimptime_2024}, who found a mean pre-processed fraction of $\sim25$ per cent (with significant scatter) for clusters with halo masses between $\sim10^{14}$\,M\textsubscript{\sun} and $\sim10^{14.5}$\,M\textsubscript{\sun}.

The pre-processed fractions shown in Figure~\ref{fig:ICL_orig_vp} feature one significant outlier: ID 13, the only cluster with a pre-processed fraction $\gtrsim30$ per cent. As shown in Table~\ref{tab:cluster_props}, this is also the most massive cluster (both in DM and in stellar mass), that with the greatest fraction of its stellar mass contained in satellites, as well as that with the lowest assembly redshift (at $z\approx0$ this cluster is uniquely part-way through two simultaneous major cluster mergers). It is also of note that the cluster with the lowest $z\approx0$ pre-processed fraction -- ID 9 -- is the least massive (both in DM and in stellar mass), that with the lowest fraction of its stellar mass contained in satellites, as well as that with the highest assembly redshift. These findings appear compatible with those of \citet{chun_formation_2023}, who previously noted a trend towards higher BCG\,+\,ICL stellar mass fractions in more relaxed clusters. 

In addition to pre-processing, we note that the category labelled as such in Figure~\ref{fig:ICL_orig_vp} also includes minor contributions from sub-threshold galaxies not detectable by the structure finder, ICL already present at high redshifts, and possibly from star formation directly into the ICL (\citealt{puchwein_intracluster_2010}, \citealt{ahvasi_insitu_2024}, and Bah\'{e} et al [in preparation]). We investigated these minor channels and find that at most $\sim0.1$ per cent of the stacked $z\approx0$ ICL of the 11 clusters could be from in-situ star formation\footnote{We estimate the in-situ fraction as the fraction of $z\approx0$ ICL stars that were first seen with an age less than the elapsed time since the previous coarse snapshot, have never (in any coarse snapshot) been seen outside their $z\approx0$ cluster (i.e outside $r_{178}$), and have never (in any coarse snapshot) been classed as part of a stellar structure by \textsc{AdaptaHOP}. Any to-be $z\approx0$ ICL stars potentially formed outside of a galaxy but beyond $r_{178}$ -- such as in a group that was later accreted by the cluster -- are considered pre-processed ICL instead. We regard this estimate as an upper limit as it may include contamination from other minor channels, such as stars that are both born and liberated into the ICL between coarse snapshots.}, and never find more than $\sim0.1$ per cent of the $z\approx0$ ICL of any cluster to have assembled before we begin monitoring (typically $\sim0.01$ per cent). Based on the analysis described in Section~\ref{ssec:hist}, we estimate less than $5$ per cent of the stacked $z\approx0$ ICL of all 11 clusters to be from unresolved galaxies with stellar masses $< 10^9$\,M\textsubscript{\sun}.

\subsection{Contribution of infalling galaxies to the ICL} \label{sec:main_results}

\subsubsection{Fraction of stars liberated from infalling galaxies} \label{ssec:strip_frac_plot}

\begin{figure}
	\includegraphics[width=\columnwidth]{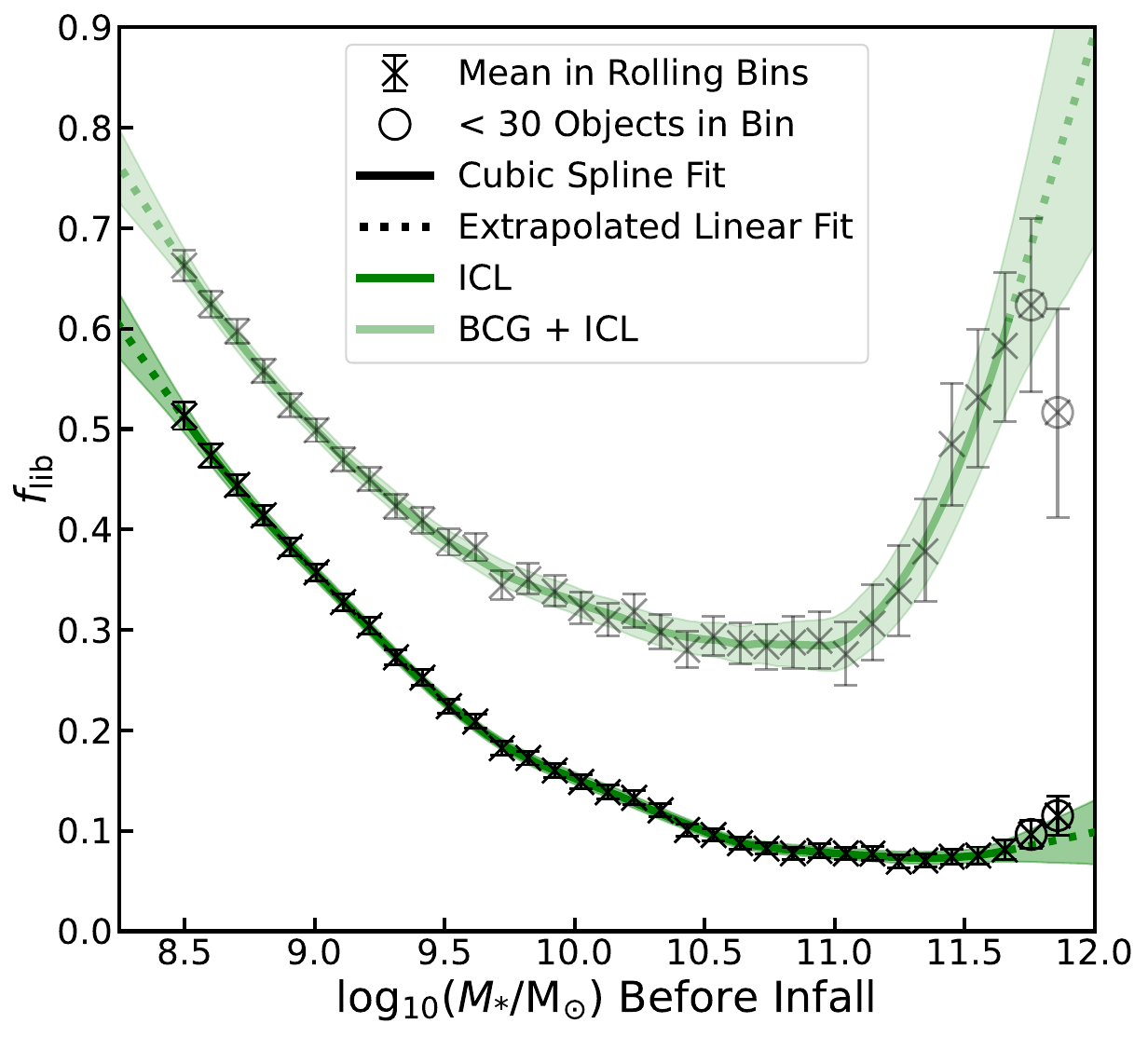}
    \caption{Mean fraction of liberated stars, $f_{\text{lib}}$, as a function of progenitor galaxy infall stellar mass, $M_{*}$, for galaxies that fell into any of the 11 clusters. Calculated using 0.5\,dex wide rolling bins. 
    A fit to the bins containing more than 30 galaxies with an added \citet{savitzky_smoothing_1964} filter is shown.
    The error bars and shaded regions indicate estimated uncertainties based on bootstrapping. The faint line and symbols show the equivalent fractions for the combined BCG\,+\,ICL system.}
    \label{fig:strip_frac_plot}
\end{figure}

We refer to the fraction of star particles tagged to the same progenitor infaller galaxy that become part of the $z\approx0$ ICL -- regardless of the mechanism that adds those stars to the ICL and considering both stars that were part of the galaxy prior to infall as well as those that later formed in a descendant of the infaller (see Section~\ref{Methods:mycode}) -- as the liberated fraction, $f_{\mathrm{lib}}$, of that progenitor. Figure~\ref{fig:strip_frac_plot} shows the mean $f_{\mathrm{lib}}$ values of tracked infaller galaxies from all 11 clusters in rolling infall stellar mass bins ($0.5$\,dex width) together with their best-fitting cubic spline fit. We use a total of 500 bins between $\mathrm{log}(M_{*}/\mathrm{M_{\sun}})\approx8.2$ and $\mathrm{log}(M_{*}/\mathrm{M_{\sun}})\approx12.1$ (the range of infall stellar masses seen in the 11 clusters) though bins containing fewer than 30 galaxies from the original sample are always ignored for fitting. Outside the infalling galaxy stellar mass range used for fitting, the fit line is linearly extrapolated. The shaded regions indicate the dispersion of fit lines generated in the same fashion for each of the bootstrap resamples. For clarity, only every fifteenth bin is shown in Figure~\ref{fig:strip_frac_plot}. The equivalent analysis for the combined BCG\,+\,ICL system is also shown in faint. 

As we expect any features in the $f_{\mathrm{lib}}$ fractions on scales smaller than a few tenths of a dex in mass to be the result of noise rather than physically meaningful features, we apply a \citet{savitzky_smoothing_1964} filter to the mean $f_{\mathrm{lib}}$ values using a 0.1\,dex wide stellar mass window before fitting. Additionally, for this analysis we ignore progenitor galaxies which only ``skimmed'' a cluster -- entering and then exiting $r_{178}$ in consecutive coarse snapshots and never returning before $z\approx0$ -- as well as those that only cross $r_{178}$ for the first time at $z\approx0$. We exclude these galaxies as we consider them distinct from the population which fall into the clusters $\gtrsim1$\,Gyr prior to $z\approx0$ and are allowed ample opportunity to be processed by the cluster.
By repeating our analysis with these galaxies included we verify the best-fitting function remains nearly unaffected, bar a small (order $0.01$) approximately uniform shift to lower values of $f_{\mathrm{lib}}$. 

From Figure~\ref{fig:strip_frac_plot} it can be seen that a less massive infalling galaxy is expected to contribute a higher fraction of its stars to the ICL. Despite the most massive galaxies being those that contribute the smallest fraction of their infall stellar mass to just the ICL alone, these galaxies are expected to make major contributions to the combined BCG\,+\,ICL system -- both in terms of absolute mass and relative to their infall stellar mass. 

As they experience stronger dynamical friction, more massive infalling galaxies should fall into the centre of a cluster faster \citep{Contini_review_2021}. Less massive galaxies, on the other hand, will have shallower potential wells and so should be more easily stripped of their stars by gravitational interactions within the cluster \citep{read_tidal_2006}. 
These two phenomena can explain the main trends seen in Figure~\ref{fig:strip_frac_plot}: very massive infalling galaxies contribute only a small fraction of their stars to the ICL as these galaxies are less easily stripped of their stars during infall, and are expected to spiral into the cluster centre (to merge with the BCG) relatively quickly. Conversely, the least massive infalling galaxies should be the most easily stripped and are expected to typically travel into the cluster centre comparatively slowly, providing ample opportunity for stripping to siphon a large fraction of their stellar mass into the ICL. This perspective is supported by a notably higher fraction of the $z\approx0$ ICL stars contributed by more massive progenitor galaxies having previously (between progenitor infall and $z\approx0$) been part of the BCG of their cluster (not shown in Figure~\ref{fig:strip_frac_plot}): $\sim25$ per cent of all $z\approx0$ ICL stars (from all 11 clusters) linked to progenitors with infall stellar masses between $10^{8.5}$\,M\textsubscript{\sun} and $10^{9.5}$\,M\textsubscript{\sun} were previously part of the BCG of their cluster, rising to $\sim37$ per cent for stars from progenitors with infall stellar masses between $10^{10.5}$\,M\textsubscript{\sun} and $10^{11.5}$\,M\textsubscript{\sun}. 

The dominant role of violent mergers as the mechanism that liberates stars from more massive galaxies into the ICL (rather than gradual stripping) is additionally supported by massive infalling galaxies that merge with the BCG before $z\approx0$ typically having significantly higher $f_{\mathrm{lib}}$ values compared with those massive infallers that survive as satellites (not shown in Figure~\ref{fig:strip_frac_plot}). 
The median (mean) value for $f_{\mathrm{lib}}$ (for the ICL alone) for all $10.5<\mathrm{log}_{10}(M_{*}/\mathrm{M_{\sun}})<11.5$ infallers from any of the 11 clusters  that did not merge with their BCG before $z\approx0$ is $\sim0.04$ ($\sim0.06$), as opposed to $\sim0.10$ ($\sim0.13$) for progenitors that did merge with the BCG. More massive infallers quickly merging with the BCG and so violent mergers becoming an increasingly important channel for adding their stars to the ICL may also explain the flattening that can be seen at the high-mass end of the fit for $f_{\mathrm{lib}}$ for the ICL alone in Figure~\ref{fig:strip_frac_plot}.


We caution that the shape of the fit for $f_{\mathrm{lib}}$ against infall stellar mass seen in Figure~\ref{fig:strip_frac_plot} for the ICL alone is suspected to be highly sensitive to the specific ICL definition employed. As described in Section~\ref{ssec:vp_plots}, different ICL definitions differ most significantly on the border between the BCG and ICL. If an alternative ICL definition were employed that moved the border between the two towards smaller cluster-centric radii, we expect the $f_{\mathrm{lib}}$ fit for the ICL system alone would move upwards to more closely resemble that seen in Figure~\ref{fig:strip_frac_plot} for the combined BCG\,+\,ICL system. Though exploring this effect of ICL definition further is beyond the scope of this study, we intend to investigate this in a future work using a broader range of simulations.

\subsubsection{Infalling galaxy mass function} \label{ssec:hist}

\begin{figure}
	\includegraphics[width=\columnwidth]{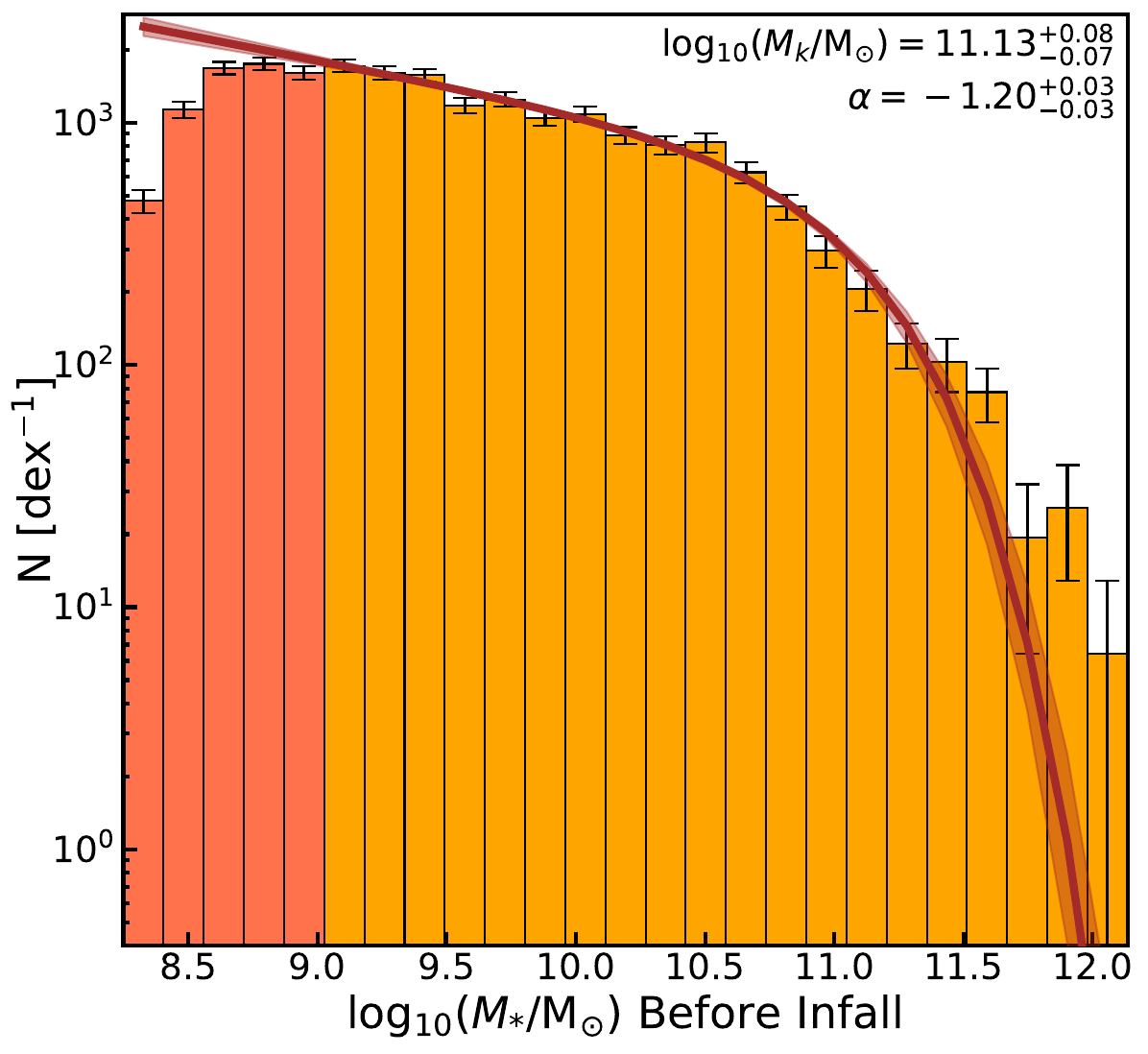}
    \caption{Stellar masses at infall, $M_{*}$, for galaxies that fall into any of the 11 clusters after $z\sim2$. A Schechter function (Equation~\ref{eq:Schechter}) is fitted using only bins with $M_{*}>10^9 M_{\sun}$; bins used for fitting are shown in orange and bins ignored for fitting are shown in red. The best-fitting parameters are presented in the upper-right and the corresponding fit line is shown in brown. The error bars and shaded regions around the fit line indicate estimated uncertainties based on bootstrapping.}
    \label{fig:hist_with_fit}
\end{figure}


Along with determining the typical contribution to the ICL made by a single galaxy infalling with a particular stellar mass, quantifying the overall expected contribution to the ICL from the entire population of infalling galaxies of a given stellar mass also requires assessing how often galaxies that massive join clusters.
A histogram of the stellar masses on infall, $M_{*}$, of galaxies which fell into any of the 11 clusters after we began to monitor ICL assembly is shown in Figure~\ref{fig:hist_with_fit}.
A fit to these data is also shown, which adopts the form of a \citet{schechter_analytic_1976} function, 
\begin{equation}
    \Phi(M_{*})\cdot dM_{*} = \Phi_{n} \left( \frac{M_{*}}{M_{k}} \right)^{\alpha} \exp\left({-M_{*} / M_{k}} \right) \cdot dM_{*}
    \label{eq:Schechter}
\end{equation}
where $\alpha$ is the low-mass-end slope of the function, $\Phi_{n}$ the normalisation, and $M_{k}$ corresponds to the ``knee'' of the function (i.e. the infall stellar mass when the function exhibits a rapid change in slope). Due to the star particle mass in \textsc{Horizon-AGN} and the minimum membership threshold of the \textsc{AdaptaHOP} structure finder, the minimum stellar mass of a detectable galaxy is $\sim10^{8}$\,M\textsubscript{\sun}. The resulting influence of resolution effects can be noted in Figure~\ref{fig:hist_with_fit} for infall stellar masses $\lesssim 10^{9}$\,M\textsubscript{\sun}. To mitigate the impact of these resolution effects on the resulting fitted function we therefore only use data from $M_{*}> 10^{9}$\,M\textsubscript{\sun} when fitting the Schechter function. 

For consistency with the analysis described in Section~\ref{ssec:strip_frac_plot}, galaxies just falling into a cluster at $z\approx0$ or those that only ``skimmed'' a cluster rather than infalling are excluded from the histogram shown in Figure~\ref{fig:hist_with_fit}. We have verified that the best-fitting parameters are nearly unchanged when these galaxies are included, so that our conclusions are unaffected by this choice. 

\subsubsection{ICL contribution as a function of infall stellar mass} \label{ssec:final_results_ssec}

\begin{figure}
	\includegraphics[width=\columnwidth]{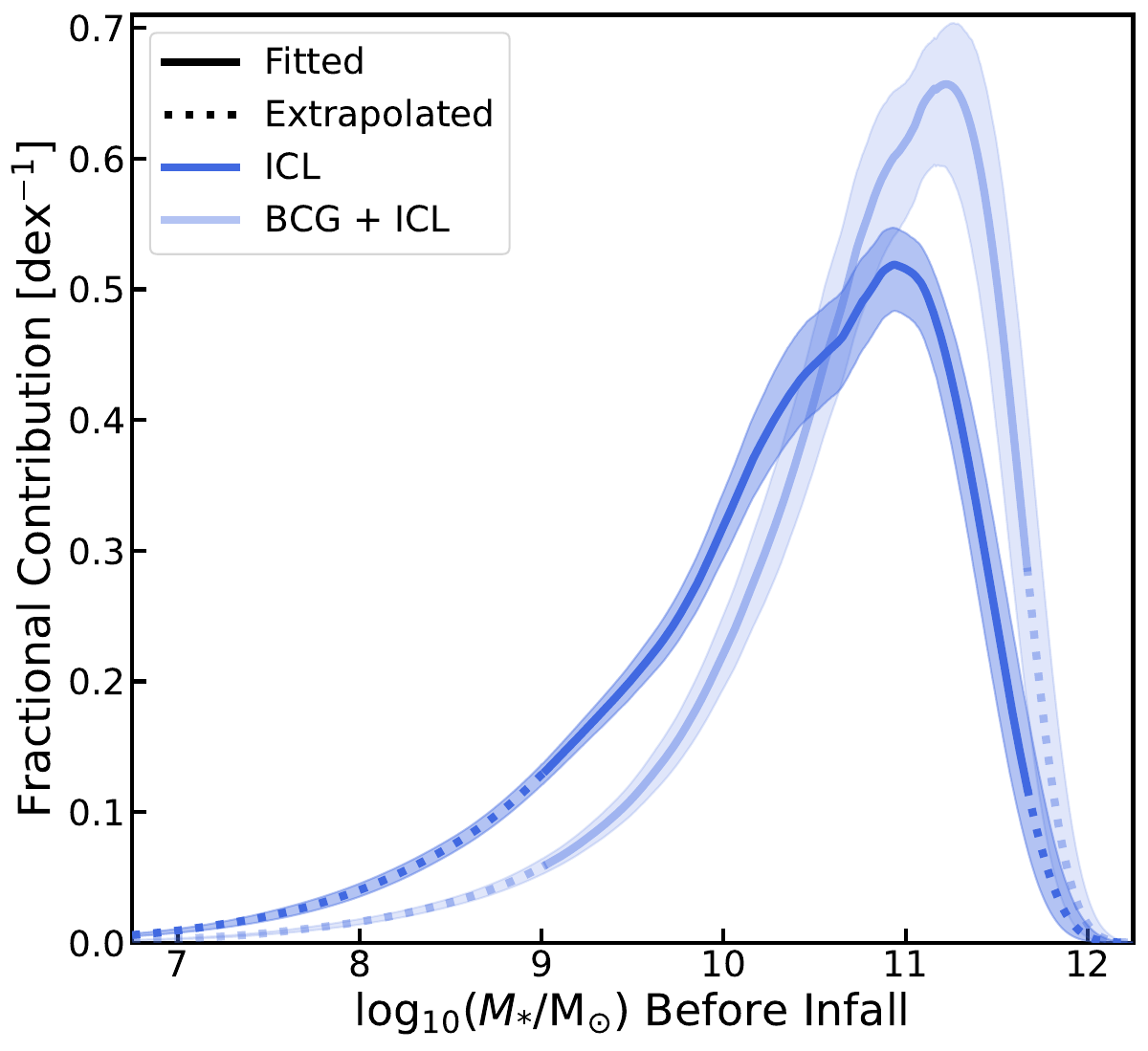}
    \caption{Fractional contribution of galaxies with different stellar masses, $M_{*}$, on cluster infall to the build-up of the (not pre-processed) ICL, stacked across all 11 clusters. Dotted lines indicate extrapolation beyond the range of masses used for fitting. The shaded regions indicate estimated uncertainties based on bootstrapping. The faint line shows the equivalent fractional contribution of infalling galaxies to the combined BCG\,+\,ICL system.}
    \label{fig:ICL_contri_plot}
\end{figure}

By combining our infall stellar mass function fit ($\Phi(M_{*})$; from Figure~\ref{fig:hist_with_fit}) with our fit for the liberated fraction as a function of infall stellar mass ($f_{\mathrm{lib}}(M_{*})$; from Figure~\ref{fig:strip_frac_plot}), and scaling by infall stellar mass ($M_{*}$),
we produce a fit for the fraction of the stacked, not pre-processed ICL of the 11 clusters associated with progenitors of a particular infall stellar mass, i.e.
\begin{equation}
    f(M_{*})=\frac{M_{*}f_{\mathrm{lib}}(M_{*})\Phi(M_{*}) }{ \int M_{*}f_{\mathrm{lib}}(M_{*})\Phi(M_{*}) \cdot dM_{*} }.
    \label{eq:final_fit}
\end{equation}
The result is shown in Figure~\ref{fig:ICL_contri_plot} (normalized to per dex in infall stellar mass), with dotted lines used to indicate extrapolation beyond the fit range of the constituent functions (which allows us to compensate for the resolution limit of the simulation). The equivalent analysis for the combined BCG\,+\,ICL system is also shown in Figure~\ref{fig:ICL_contri_plot} with a fainter blue line. 

We find no clear trend (and considerable scatter) between the infall stellar mass of each galaxy and the fraction of the stars associated with that infaller that were formed during or after cluster infall. We therefore assume a constant ratio between the total stellar mass associated with a progenitor at $z\approx0$ and its infall stellar mass, and consequently neglect post-infall star formation in Equation~\ref{eq:final_fit} as this constant ratio has no impact on the shape of the curve seen in Figure~\ref{fig:ICL_contri_plot}. The median (mean) ratio between the total associated stellar mass at $z\approx0$ and the infall stellar mass of a progenitor for all $\sim4000$ infaller galaxies across all 11 clusters was $\sim1.01$ ($\sim1.18$).

It can be seen in Figure~\ref{fig:ICL_contri_plot} that the dominant contributors of stars to the ICL are galaxies with infall stellar masses between $\sim10^{10.5}$\,M\textsubscript{\sun} and $\sim10^{11}$\,M\textsubscript{\sun}. This is in good agreement with the general conclusion of most prior theoretical studies (e.g. \citealt{contini_formation_2014, contini_theoretical_2019, chun_formation_2023, chun_formation_2024, ahvazi_progenitors_2024}) -- that approximately Milky-Way mass galaxies are the main progenitors of the ICL. 

We highlight that the location of the peak in Figure~\ref{fig:ICL_contri_plot} -- at $\log_{10}(M_{*}/\textrm{M}_{\sun}) = 10.94^{+0.13}_{-0.07}$ (uncertainty estimated by bootstrapping) -- is very close to the location of the ``knee'' of the Schechter function fitted in Figure~\ref{fig:hist_with_fit} ($\log_{10}(M_{k}/\textrm{M}_{\sun}) = 11.13^{+0.08}_{-0.07}$). Additionally, we note that the position of the peak in Figure~\ref{fig:ICL_contri_plot} varies very little between the main version of our analysis and the alternative version in which the BCG and ICL are not separated (peak at $\log_{10}(M_{*}/\textrm{M}_{\sun}) = 11.2^{+0.1}_{-0.3}$), despite the two fits used for $f_{\text{lib}}$ as a function of infall mass being significantly different (as can be seen in Figure~\ref{fig:strip_frac_plot}). This is a consequence of both these fits for $f_{\text{lib}}$ only varying by a factor of order unity over a nearly four orders of magnitude change in infall stellar mass, with the fitted infall stellar mass function (shown in Figure~\ref{fig:hist_with_fit}) instead varying by several orders of magnitude over this same infall mass range. As such, the specific fit used for $f_{\text{lib}}$ as a function of infall mass has only a minor effect on the shape of the curve seen in Figure~\ref{fig:ICL_contri_plot}, which is instead governed almost entirely by the infall stellar mass function. 

Though Figure~\ref{fig:ICL_contri_plot} shows that the expected contribution of infalling objects with $M_{*} \gtrsim 10^{11}$\,M\textsubscript{\sun} to the ICL rapidly declines with any further increase in mass, this is entirely due to how rarely we expect such supremely massive objects to fall into clusters (as can be noted from Figure~\ref{fig:hist_with_fit}); much of the mass of such a massive galaxy would need to be assembled through mergers, and so can only be assembled at the heart of a rich group or cluster, hence such objects only join clusters very rarely (as part of major cluster mergers). However, if such a massive object does join a cluster we anticipate that it would produce a significant contribution to the ICL of that cluster, even before considering the pre-processed ICL that should accompany it. This view agrees with the findings of \citet{cooper_surface_2015} and \citet{harris_quantifying_2017}: that the ICL of a cluster can be heavily influenced by only a small number of massive progenitors, being largely built up stochastically as these rare massive objects infrequently join the cluster. This stochasticity may also explain the large dispersion found in the fractions of $z\approx0$ ICL stars that were formerly part of the BCG (shown in Figure~\ref{fig:ICL_orig_vp}), as violent mergers between the BCG and the most massive infallers are expected to be the primary mechanism for liberating BCG stars into the ICL.

It warrants restating that the analysis depicted in Figures \ref{fig:strip_frac_plot} and \ref{fig:ICL_contri_plot} ignores pre-processed ICL. If this analysis were repeated with the pre-processed ICL from accreted groups/clusters instead attributed to the central galaxies of those groups/clusters, we would expect the shapes of the fits seen in Figures \ref{fig:strip_frac_plot} and \ref{fig:ICL_contri_plot} to be substantially altered, with this extra attributed ICL mass enhancing the fractional contribution of infalling galaxies with $M_{*}\gtrsim10^{11}$\,M\textsubscript{$\sun$} (as well as the $f_{\textrm{lib}}$ values determined for these galaxies), potentially shifting the location of the peak seen in Figure~\ref{fig:ICL_contri_plot} to a slightly higher mass. 

\begin{figure}
	\includegraphics[width=\columnwidth]{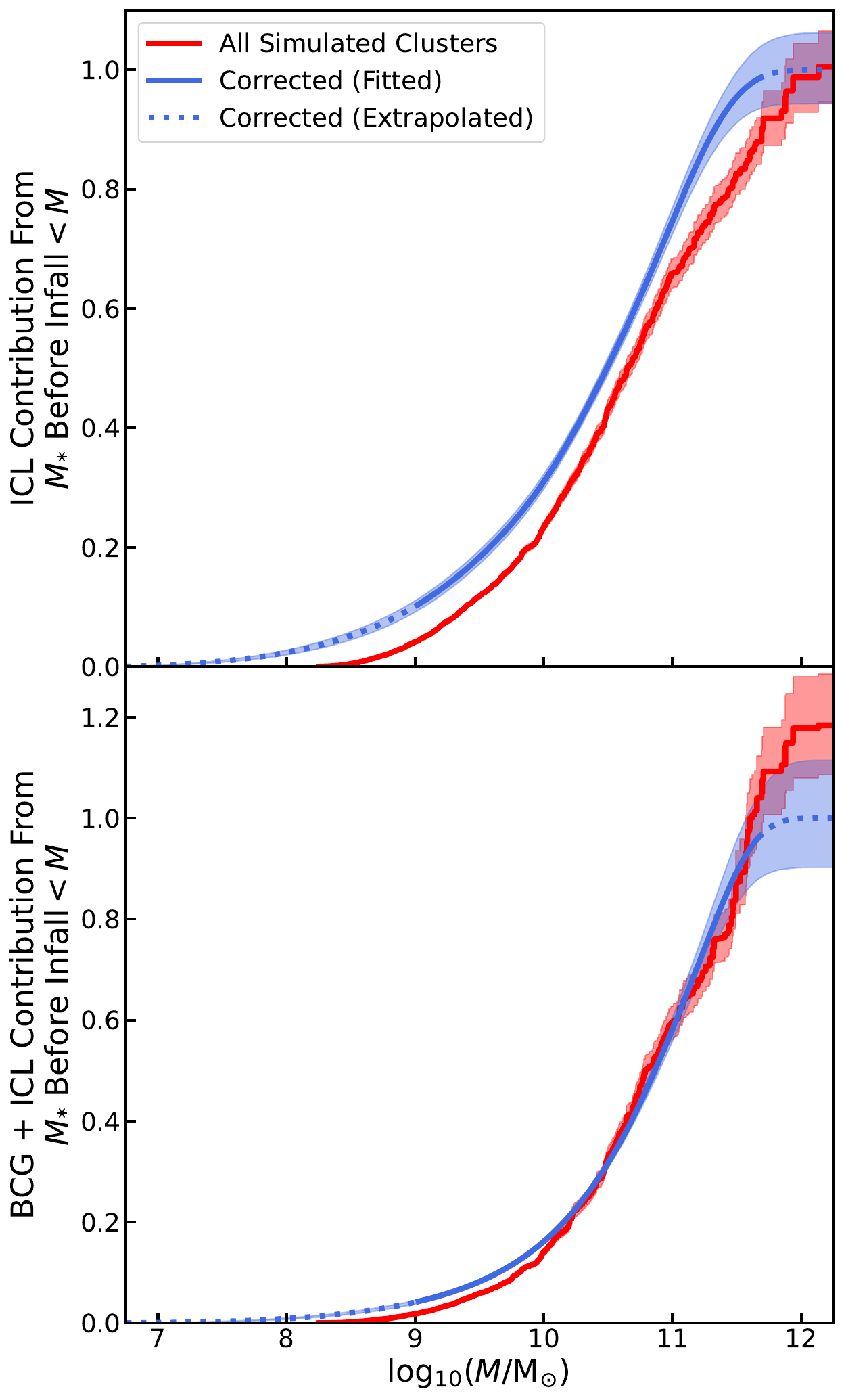}
    \caption{\textbf{Top:} Normalised cumulative contribution to the stacked (not pre-processed) ICL of the 11 clusters by galaxies with stellar mass, $M_{*}$, on cluster infall less than $M$ after correcting for resolution effects (shown in blue). Dotted lines are used to indicate extrapolation beyond the range of masses used for fitting. Shown in red (for comparison) is a cumulative sum plot for the stacked (not pre-processed) ICL taken directly from the 11 simulated clusters (with no corrections made for resolution effects), normalized relative to the corrected curve. The shaded regions indicate estimated uncertainties based on bootstrapping. \textbf{Bottom:} Same as the top panel but for the combined BCG\,+\,ICL system. }
    \label{fig:scaled_CS_plot}
\end{figure} 

The top panel of Figure~\ref{fig:scaled_CS_plot} shows the (normalised) cumulative contribution to the stacked, not pre-processed ICL of all 11 clusters from progenitor objects with infall stellar masses, $M_{*}$, less than $M$: the red curve is derived directly from the simulation (without correcting for resolution effects), and the blue curve is derived from the fitted function depicted in Figure~\ref{fig:ICL_contri_plot}. Dashed lines are once again used to indicate extrapolation beyond the range of infall stellar masses used for fitting. The corrected curve is normalised to unity, and the uncorrected curve is normalised relative to the corrected curve (adjusted by a constant factor equal to the mean ratio between total associated stellar mass at $z\approx0$ and progenitor infall stellar mass to account for star formation during and after cluster infall), so that the total ICL mass encapsulated by both curves can be directly compared. The bottom panel of Figure~\ref{fig:scaled_CS_plot} shows the same as the top panel but for the combined BCG\,+\,ICL system instead. 

It can be seen in the top panel of Figure~\ref{fig:scaled_CS_plot} that $\sim50$ per cent of the (not pre-processed) ICL is predicted to come from galaxies with infall stellar masses within half a dex of $10^{11}$\,M\textsubscript{\sun}, and $\sim90$ cent is predicted to come from progenitors with stellar masses $\gtrsim10^9$\,M\textsubscript{\sun}. Even after compensating for resolution effects using our extrapolated fitting function approach, that galaxies with $M_{*}<10^{9}$\,M\textsubscript{\sun} contribute only $\lesssim10$ per cent of the total ICL mass suggests that the bulk properties of the ICL should be insensitive to contributions from these low-mass galaxies. This also tentatively implies that once a simulation achieves sufficient resolutions to resolve galaxies with $M_{*}\sim10^{9}$\,M\textsubscript{\sun}, further improvements in resolution will yield no significant changes in the bulk properties of the ICL and only minor changes ($\lesssim10$ per cent) in the total ICL masses found. 

When \cite{puchwein_intracluster_2010} previously investigated the effects of altering the mass resolution of a hydrodynamical simulation on the ICL masses identified in simulated clusters, beyond a star-particle mass of $\sim10^{7}$\,M\textsubscript{\sun} the ICL masses of their simulated clusters appear to have converged -- a suggestion which supports our findings here. Likewise, our findings agree well with those of \cite{ahvazi_progenitors_2024}, who found that $\gtrsim90$ per cent of the ICL in the groups and clusters from the higher resolution \textsc{TNG50} simulation that they investigated typically came from progenitor objects with stellar masses greater than $\sim10^9$\,M\textsubscript{\sun} and that $\gtrsim50$ per cent typically came from progenitors with stellar masses greater than $\sim10^{10}$\,M\textsubscript{\sun} (though with significant dispersion). 

It is worth noting that the shaded regions in the Figure~\ref{fig:scaled_CS_plot} (indicating estimated uncertainties based on bootstrapping) only significantly broaden for stellar masses beyond $\sim10^{11}$\,M\textsubscript{\sun}, indicating that the substitution of even only a very small number of progenitor objects this massive for less massive ones (or vice versa) can appreciably alter the total ICL mass of all 11 clusters combined -- emphasizing the significant ICL contributions objects this massive can make when they do join clusters. 
We do note, however, that the seeming strong agreement between the corrected and uncorrected curves at the high mass end in the top panel of Figure~\ref{fig:scaled_CS_plot} is somewhat coincidental: the Schechter function fitted in Figure~\ref{fig:hist_with_fit} slightly exaggerates the scarcity of the most massive objects, with this coincidentally diminishing the expected overall contribution to the ICL from high mass infallers by approximately as much as unresolved low mass galaxies are expected to contribute. The mass function fit slightly exaggerating the scarcity of high mass infallers is also why the corrected curve in the bottom panel of Figure~\ref{fig:scaled_CS_plot} falls slightly below the uncorrected curve at the high mass end.

We emphasize that the analysis represented in Figure~\ref{fig:scaled_CS_plot} is for the combined ICL / BCG\,+\,ICL of all 11 clusters stacked together. We warn that if this analysis were repeated on any individual cluster the stochastic nature of cluster assembly could cause the shape of the resulting curve to significantly diverge beyond what is encompassed by the shaded regions in Figure~\ref{fig:scaled_CS_plot}, particularly for very high mass progenitor objects (scarce few of which should fall into such a $10^{14}$\,M\textsubscript{\sun} cluster and so the entire contribution of the high infall mass regime may come from only one or two objects). Equivalent curves for each of the 11 clusters individually are presented in Appendix~\ref{append:indiv_cs_curves}.

\section{Conclusions}\label{conclusions}

Using the \textsc{Horizon-AGN} simulation \citep{dubois_dancing_2014}, we have studied the assembly of the intracluster light (ICL). We investigated ICL assembly in 11 simulated clusters with $z\approx0$ dark matter halo masses between $\sim1\times10^{14}$\,M\textsubscript{\sun} and $\sim7\times10^{14}$\,M\textsubscript{\sun}, tracking the stars of galaxies that fell into these clusters over the past $\sim10$\,Gyr in order to quantify the differing contributions made to the $z\approx0$ ICL by galaxies with differing stellar masses on cluster infall. Our main findings can be summarised as follows:

\begin{enumerate}

    \item \noindent \textit{On average, $44$ per cent of the $z\approx0$ ICL stars in the 11 studied clusters were previously part of a satellite galaxy within the cluster halo (but never part of the BCG); another $36$ per cent on average were formerly part of the BCG} (Figure~\ref{fig:ICL_orig_vp}). Such significant fractions in both instances indicate both satellite stripping and violent BCG mergers play significant roles in ICL assembly. \\

    \item \noindent \textit{In-situ formation of stars directly into the ICL appears insignificant in the \textsc{Horizon-AGN} simulation; we discern an upper limit of order $0.1$ per cent on the fraction of the stacked ICL of the 11 clusters formed by this channel. The remainder of the ICL is virtually all pre-processed}. The average fraction of $z\approx0$ ICL stars that were pre-processed and never part of a cluster galaxy was $16$ per cent (Figure~\ref{fig:ICL_orig_vp}), though in the least relaxed and most massive cluster $\gtrsim40$ per cent of $z\approx0$ ICL stars fell into this category. Inversely, the lowest pre-processed fraction was seen in the most relaxed and least massive cluster. \\

    \item \noindent \textit{A galaxy with less stellar mass on cluster infall can be expected to contribute a greater percentage of its stars to the ICL than a more massive infalling galaxy}. The most massive galaxies that join clusters do, however, typically contribute a very high fraction of their infall stellar mass to the combined BCG\,+\,ICL system ($\gtrsim60$ per cent; Figure~\ref{fig:strip_frac_plot}).\\

    \item \noindent \textit{Roughly half of the stacked, not pre-processed ICL of the 11 clusters came from progenitor objects with infall stellar masses within half a dex of} $10^{11}$\,M\textsubscript{\sun} (Figure~\ref{fig:scaled_CS_plot}). This is the case even after compensating for resolution effects. As the mean fraction of stars liberated from infalling galaxies only varied by a factor of order unity between infall masses of $\sim10^{9}$\,M\textsubscript{\sun} and $\sim10^{12}$\,M\textsubscript{\sun}, the location of this peak is largely determined by the infalling galaxy stellar mass function instead. Any object as or more massive than the Milky-Way that joins a cluster is expected to make a sizeable contribution to the ICL of that cluster and -- as such massive objects joining clusters should be infrequent events -- this suggests ICL assembly may be chiefly stochastic. \\

    \item \noindent \textit{$90$ per cent of the bulk ICL of the 11 clusters which was not pre-processed could be attributed to progenitors with infall stellar masses} $\gtrsim10^{9}$\,M\textsubscript{\sun}, even after compensating for resolution effects (Figure~\ref{fig:scaled_CS_plot}). As the ICL appears virtually complete even without any contribution from less massive galaxies (at least within the validity of our stellar mass function extrapolation), we expect the bulk properties of the ICL to be insensitive to the low-mass galaxy population. \\

\end{enumerate}

In summary, we have shown that the main progenitors of the ICL should be massive galaxies, and more specifically those with stellar masses on cluster infall close to that of the Milky-Way. Despite less massive galaxies being more numerous and adding a larger fraction of their stellar mass on cluster infall to the ICL, the overall contribution of this population to the ICL of a $\sim10^{14}$\,M\textsubscript{\sun} halo mass cluster appears rather small. Conversely, though very massive galaxies join clusters only rarely, they make major contributions to the mass of the ICL when they do so, despite this conferred ICL mass only being a small fraction of their infall stellar mass. 

Though we do not expect the low-mass galaxy population to meaningfully influence the bulk properties of the ICL, the potential remains for this population to have a significant impact on the radial dependence of ICL properties. We expect much of the mass of more massive infalling galaxies to be deposited close to the centre of the cluster (as implied in Figure~\ref{fig:strip_frac_plot}), potentially allowing the ICL in the outskirts of a cluster to be dominated by stars from less massive progenitors - an idea supported observationally by prior studies noting metallicity and colour gradients in the ICL (e.g. \citealt{demaio_lost_2018}, \citealt{gu_spectroscopic_2020}, and \citealt{golden-marx_characterizing_2023}). We intend to investigate this further in a future work.  

\section*{Acknowledgements}

We thank the anonymous referee for their careful reading of our original manuscript and for their constructive comments which have helped to improve the quality and clarity of the presented work.
We also thank Jesse B. Golden-Marx for helpful discussions and comments. 
This work made use of \textsc{NumPy} \citep{harris_array_2020}, \textsc{Matplotlib} \citep{hunter_matplotlib_2007}, \textsc{SciPy} \citep{virtanen_scipy_2020}, 
\textsc{h5py} \citep{collette_h5pyh5py_2023}, \textsc{Astropy} \citep{the_astropy_collaboration_astropy_2022}, and \textsc{pyGAM} \citep{pyGAM_ref}.
H.~J.~Brown gratefully acknowledges support from the UK Science and Technology Facilities Council (STFC) under grant ST/Y509437/1. G.~Martin, F.~R.~Pearce, and N.~A.~Hatch gratefully acknowledge support from the UK STFC under grant ST/X000982/1. N.~A.~Hatch gratefully acknowledges support from the Leverhulme Trust through a Research Leadership Award. Y.~M.~Bah\'{e} gratefully acknowledges financial support from the Swiss National Science Foundation (SNSF) under funding reference 200021\_213076. 
This work was granted access to the HPC resources of CINES under allocations 2013047012, 2014047012 and 2015047012 made by GENCI. This work has made use of the Infinity cluster, hosted by the Institut d'Astrophysique de Paris. We warmly thank S.~Rouberol for running it smoothly.

\section*{Data Availability}

The raw data products of the \textsc{Horizon-AGN} simulation are available upon reasonable request through the collaboration’s website: \url{https://www.horizon-simulation.org/}. The data products generated from this work are available upon reasonable request from the corresponding author. 


\bibliographystyle{mnras}
\bibliography{refs} 



\appendix

\section{Individual cluster cumulative ICL contribution curves} \label{append:indiv_cs_curves}

\begin{figure}
	\includegraphics[width=\columnwidth]{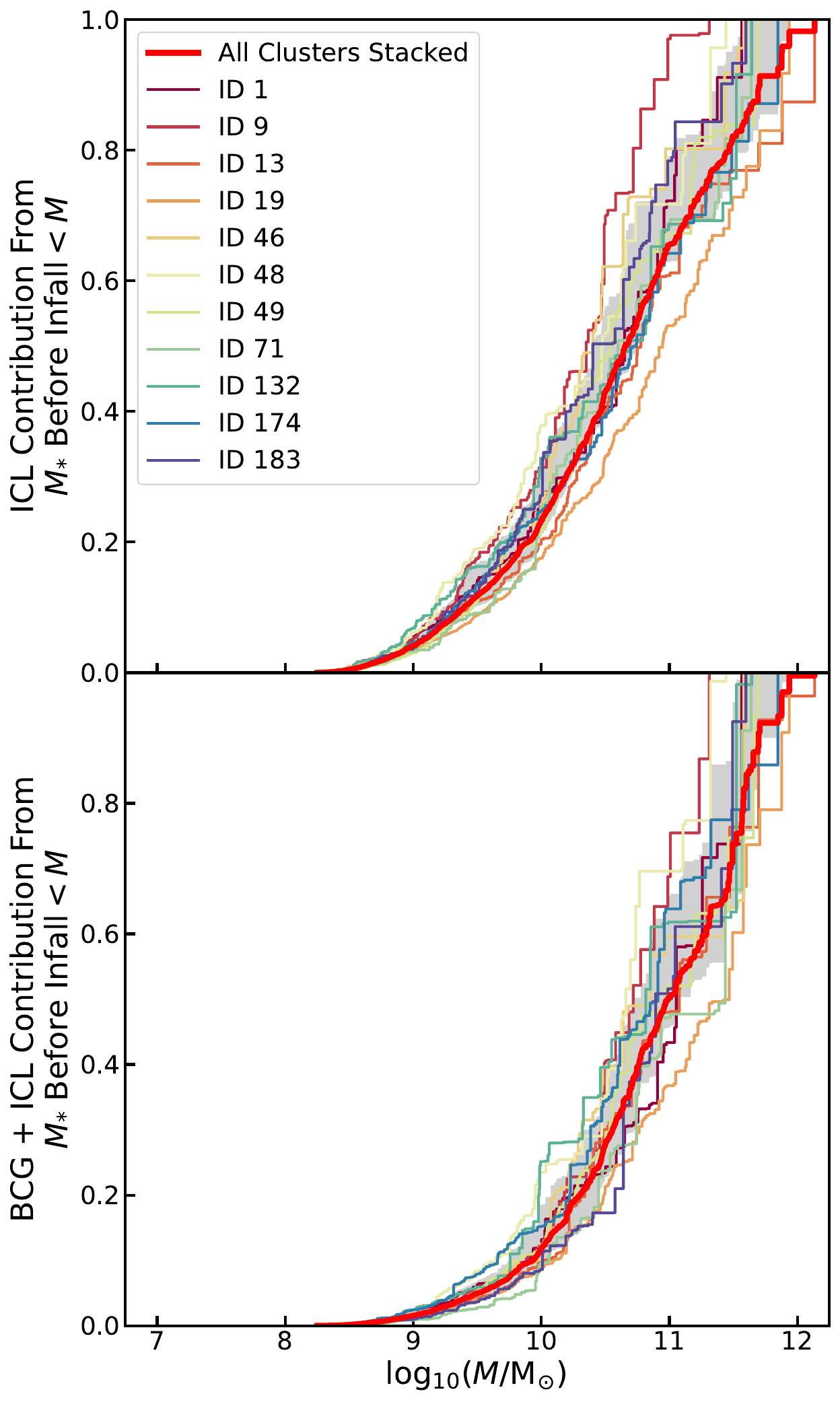}
    
    \caption{\textbf{Top:} Normalized cumulative contribution to the (not pre-processed) $z\approx0$ ICL of each of the 11 clusters by galaxies with stellar mass, $M_{*}$, on cluster infall less than $M$ (with no corrections made for resolution effects). The grey shaded region is bounded by the 16\textsuperscript{th} and 84\textsuperscript{th} percentiles of the individual cluster curves. The corresponding curve for the ICL of all 11 clusters stacked is also shown (as the thicker red line). \textbf{Bottom:} Same as the top panel but for the combined BCG\,+\,ICL system.}
    \label{fig:indiv_clus_CS_plots}
\end{figure} 

Figure~\ref{fig:scaled_CS_plot} shows the cumulative contribution to the ICL / BCG\,+\,ICL from increasingly massive progenitors for the stacked ICL / BCG\,+\,ICL of all 11 clusters combined. The uncertainties included in that plot (estimated based on bootstrapping and indicated by the shaded areas) are thus not thought to be representative of the anticipated cluster-to-cluster scatter should an equivalent analysis be performed on individual clusters. We present this equivalent analysis for each of the 11 clusters individually in Figure~\ref{fig:indiv_clus_CS_plots}, showing for each cluster the fraction of the (not pre-processed) $z\approx0$ ICL mass (top panel) or BCG\,+\,ICL mass (bottom panel) linked to progenitors with infall stellar mass, $M_{*}$, less than $M$ (taken directly from the simulation with no corrections made for resolution effects). The grey shaded regions are bounded by the 16\textsuperscript{th} and 84\textsuperscript{th} percentiles of the individual cluster curves. To facilitate comparison, the corresponding (uncorrected) curves for the ICL / BCG\,+\,ICL of all 11 clusters stacked from Figure~\ref{fig:scaled_CS_plot} (normalised to unity) are also shown. The ID numbers referenced are those presented in Table~\ref{tab:cluster_props}. 

In the top panel of Figure~\ref{fig:indiv_clus_CS_plots} the individual cluster curves for IDs 9 and 19 are worthy of particular note. Only $\sim30$ per cent of the total $z\approx0$ ICL mass of ID 9 is from progenitors with $M_{*}\gtrsim10^{10.5}$\,M\textsubscript{\sun}, and so the ICL of this specific cluster is primarily sourced from lower mass progenitors. Conversely, $\sim50$ per cent of the total $z\approx0$ ICL mass of ID 19 is from progenitors with $M_{*}\gtrsim10^{11}$ hence the ICL of this particular cluster is dominated by contributions from infalling galaxies more massive than the Milky Way. These two clusters serve as prime examples of how the main progenitors of ICL stars in individual clusters can diverge from what is expected from considering an ensemble of clusters due to the stochastic nature of cluster assembly. Several instances can also be seen in the top panel of Figure~\ref{fig:indiv_clus_CS_plots} of $\gtrsim15$ per cent of the (not pre-processed) ICL mass of an individual cluster being linked to a single massive ($M_{*}\gtrsim10^{11}$\,M\textsubscript{\sun}) progenitor. This supports the perspective that -- although they generally join clusters too infrequently to be the dominate overall contributors of ICL stars -- individual massive galaxies can potentially make major ICL contributions when they do (rarely) join clusters. 


\bsp	
\label{lastpage}
\end{document}